\documentclass[12pt]{article}
\pdfoutput=1
\usepackage[dvips]{graphicx}
\usepackage{epsfig,cite}
\usepackage {amsmath}
\usepackage{color}
\usepackage{amssymb}
\usepackage{slashed}
\include{epsf}
\newcount\timecount
\newcount\hours \newcount\minutes  \newcount\temp \newcount\pmhours
\hours = \time
\divide\hours by 60
\temp = \hours
\multiply\temp by 60
\minutes = \time
\advance\minutes by -\temp
\def\hour{\the\hours}
\def\minute{\ifnum\minutes<10 0\the\minutes
            \else\the\minutes\fi}
\def\clock{
\ifnum\hours=0 12:\minute\ AM
\else\ifnum\hours<12 \hour:\minute\ AM
      \else\ifnum\hours=12 12:\minute\ PM
            \else\ifnum\hours>12
                 \pmhours=\hours
                 \advance\pmhours by -12
                 \the\pmhours:\minute\ PM
                 \fi
            \fi
      \fi
\fi
}

\def\monthname{\relax\ifcase\month 0/\or January\or February\or
   March\or April\or May\or June\or July\or August\or September\or
   October\or November\or December\else\number\month/\fi}

\def\bold#1{\setbox0=\hbox{$#1$}%
     \kern-.025em\copy0\kern-\wd0
     \kern.05em\copy0\kern-\wd0
     \kern-.025em\raise.0433em\box0 }

\def\beq{\begin{equation}}
\def\eeq{\end{equation}}

\def\ga{\mathrel{\raise.3ex\hbox{$>$\kern-.75em\lower1ex\hbox{$\sim$}}}}
\def\la{\mathrel{\raise.3ex\hbox{$<$\kern-.75em\lower1ex\hbox{$\sim$}}}}
\def\gev{{\rm \, Ge\kern-0.125em V}}
\def\tev{{\rm \, Te\kern-0.125em V}}
\def\gyr{{\rm \, G\kern-0.125em yr}}

\def\gappeq{\mathrel{\rlap {\raise.5ex\hbox{$>$}}
{\lower.5ex\hbox{$\sim$}}}}
\def\lappeq{\mathrel{\rlap{\raise.5ex\hbox{$<$}}
{\lower.5ex\hbox{$\sim$}}}}
\def\Toprel#1\over#2{\mathrel{\mathop{#2}\limits^{#1}}}

\def\m12{m_{1\!/2}}

\def\bea{\begin{eqnarray}}
\def\eea{\end{eqnarray}}

\def\beqar{\begin{eqnarray}}
\def\eeqar{\end{eqnarray}}

\begin{document}

\begin{titlepage}
\pagestyle{empty}
\baselineskip=21pt
\rightline{KCL-PH-TH/2013-47, LCTS/2013-35, CERN-PH-TH/2013-312}
\vskip 0.8in
\begin{center}
{\large {\bf Disentangling Higgs-Top Couplings in Associated Production}}

\end{center}
\begin{center}
\vskip 0.4in
 {\bf John~Ellis}$^{1,2}$,
{\bf Dae~Sung~Hwang}$^{3}$,
{\bf Kazuki~Sakurai}$^1$
and {\bf Michihisa~Takeuchi}$^1$
\vskip 0.1in
{\small {\it
$^1${Theoretical Particle Physics and Cosmology Group, Physics Department, \\
King's College London, London WC2R 2LS, UK}\\
$^2${TH Division, Physics Department, CERN, CH-1211 Geneva 23, Switzerland}\\
$^3${Department of Physics, Sejong University, Seoul 143-747, South Korea}\\
}}
\vskip 0.6in
{\bf Abstract}
\end{center}
\baselineskip=18pt \noindent
{
In the presence of CP violation, the Higgs-top coupling may have both scalar and pseudoscalar components, $\kappa_t$ and ${\tilde \kappa_t}$,
which are bounded indirectly but only weakly by the present experimental constraints
on the Higgs-gluon-gluon and Higgs-$\gamma$-$\gamma$ couplings, whereas upper limits on electric dipole
moments provide strong additional indirect constraints on ${\tilde \kappa_t}$, if the Higgs-electron coupling is similar
to that in the Standard Model and there are no cancellations with other contributions. We discuss
methods to measure directly the scalar and pseudoscalar Higgs-top couplings by measurements of Higgs
production in association with ${\bar t} t$, single $t$ and single ${\bar t}$ at the LHC.
Measurements of the total cross sections are very sensitive to variations in the Higgs-top couplings
that are consistent with the present indirect constraints, as are invariant mass distributions in
${\bar t} t H$, $tH$ and ${\bar t} H$ final states. We also investigate the additional
information on $\kappa_t$ and ${\tilde \kappa_t}$ that could be obtained from measurements of the longitudinal and transverse
$t$ polarization in the different associated production channels, and the ${\bar t} t$ spin
correlation in ${\bar t} t H$ events.}


\vfill
\leftline{
December 2013}
\end{titlepage}
\baselineskip=18pt

\section{Introduction}

It is important to characterize the new boson $H$ discovered by the ATLAS~\cite{Aad:2012tfa} and CMS~\cite{Chatrchyan:2012ufa} Collaborations. In this paper we refer to this particle as a Higgs boson, since it has some of the expected properties,
though others remain to be verified. Tests via $H$ decays into $\gamma \gamma$~\cite{atlas-gamgam-spin}, 
$W W^*$~\cite{atlas-WW-spin} and $Z Z^*$~\cite{atlas-ZZ-spin, cms-ZZ-spin} 
are consistent with it having spin zero~\cite{Bolognesi:2012mm, Ellis:2012mj, EHspin, Freitas:2012kw, atlas-comb-spin}, as are
measurements of $H$ production in association with $W$ and $Z$~\cite{Ellis:2013ywa}. 
In particular, these tests
exclude graviton-like spin-two couplings with a high degree of confidence.
Assuming that the $H$ spin
is indeed zero, the next question is whether it has scalar and/or pseudoscalar couplings. 

Tests in $W W^*$ and $Z Z^*$ final states~\cite{Godbole:2007cn, Gao:2010qx, Bolognesi:2012mm, Coleppa:2012eh, Freitas:2012kw, Djouadi:2013qya, atlas-ZZ-spin, cms-ZZ-spin, Stolarski:2012ps} and production in association with $W$ and $Z$~\cite{Ellis:2013ywa} also
disfavour strongly pure pseudoscalar couplings, but do not yet exclude a substantial pseudoscalar
admixture.  In the presence of CP violation, the ratios of scalar and pseudoscalar couplings may 
differ from channel to channel, and it is important to measure them in as many different channels
as possible. 
Strategies to measure a CP-violating admixture in $H \to \tau^+ \tau^-$ decays have
been proposed~\cite{Plehn:2001nj, Harnik:2013aja, Berge:2011ij}, and other tests are possible in $H$ production 
in vector-boson fusion~\cite{Hagiwara:2009wt, Englert:2012ct, Andersen:2012kn, Englert:2012xt, Englert:2013opa},
 double-diffractive~\cite{Khoze:2004rc, Ellis:2005fp} and $\gamma \gamma$ collisions~\cite{Badelek:2001xb}.

There are already indirect constraints on the scalar and pseudoscalar $H$-top couplings
$\kappa_t$ and ${\tilde \kappa_t}$ from experimental information on the $H$-gluon-gluon and 
$H$-$\gamma$-$\gamma$ couplings~\cite{Brod:2013cka}, but these constraints are relatively weak~\cite{Ellis:2013lra}, 
as we discuss later. 
Upper limits on electric dipole moments also impose important indirect constraints on a possible 
pseudoscalar $H$-top coupling ${\tilde \kappa_t}$~\cite{Brod:2013cka}, if one assumes that 
if the Higgs-electron coupling is similar to that in the Standard Model and there
are no cancellations with other contributing mechanisms. 

In this paper we investigate the potential for 
disentangling scalar and pseudoscalar $H$-top couplings directly at the LHC,
using measurements of $H$ production in association with ${\bar t} t$, single $t$ and single ${\bar t}$.
These processes offer many observables that can contribute to determining $\kappa_t$ and ${\tilde \kappa_t}$,
including the total cross sections, ${\bar t} t H$, $t H$ and ${\bar t} H$ invariant mass distributions and
various $t ({\bar t})$ polarization observables. These include polarizations both within and
perpendicular to the production plane. The latter are particularly interesting, since they violate
CP explicitly.

The search strategy for the ${\bar t} t H$ process has been studied in various Higgs decay modes:
$\bar bb$~\cite{Plehn:2009rk, Buckley:2013auc}, 
$\tau\tau$~\cite{Boddy:2012nt} and $WW^*$~\cite{Maltoni:2002jr}.
ATLAS~\cite{TheATLAScollaboration:2013mia} and CMS~\cite{CMS:2013fda} 
have searched for this process intensively using the 8~TeV data set, but the current luminosity and analyses have not reached the 
sensitivity required by the Standard Model Higgs boson.
The associated production of the Higgs and a single top has recently attracted attention since there is a large destructive interference between two Feynman diagrams 
with $\bar t t H$ and $WWH$ couplings in the Standard Model and the production cross section is sensitive to the deviation of the couplings from the
the Standard Model values.   
The dependences on these couplings of the cross section and Higgs branching ratios 
as well as the search strategy have been studied in~\cite{Maltoni:2001hu, Farina:2012xp}
assuming CP-conserving interactions.


The structure of this paper is as follows. In Section~2 we introduce the scalar and pseudoscalar 
$H$-top couplings $\kappa_t$ and ${\tilde \kappa_t}$ and discuss the current indirect experimental
constraints, paying particular attention to those provided by LHC constraints on the $H$-gluon-gluon and 
$H$-$\gamma$-$\gamma$ couplings, taking their correlations into account~\cite{Ellis:2013lra}. Section~3 presents
calculations of the total cross sections for $H$ production in association with ${\bar t} t$,
single $t$ and single ${\bar t}$. We show that, within the region of the $(\kappa_t, {\tilde \kappa_t})$
plane allowed at the 68\% CL, the total cross section for ${\bar t} t H$ production may be
considerably {\it smaller} than in the Standard Model, whereas the cross sections for $t H$ and ${\bar t} H$
may be considerably {\it larger}. As we show in Section~4, the ${\bar t} t H$, $t H$ and ${\bar t} H$
invariant mass distributions may also be very different from those expected in the Standard Model.
We proceed in Section~5 to discuss the possibilities for $t$ polarization measurements at the LHC.

Our results indicate that the LHC operating at 13/14~TeV may soon be able to
provide interesting direct constraints on $\kappa_t$ and ${\tilde \kappa_t}$, including direct constraints
on CP violation in the top sector.

\section{Indirect Constraints on Top-Higgs Couplings}

We write the top-$H$ couplings in the form
\begin{equation}
{\cal L}_t \; = \; - \frac{m_t}{v} \left( \kappa_t {\bar t}t + i {\tilde \kappa}_t {\bar t} \gamma_5 t \right) H \, ,
\label{kappas}
\end{equation}
where $v = 246$~GeV is the conventional Higgs vacuum expectation value (v.e.v.)
and $\kappa_t = 1$ and ${\tilde \kappa}_t = 0$ in the Standard Model.

As observed in~\cite{Brod:2013cka}, the ${\tilde \kappa}_t$ coupling makes an important contribution to the
electric dipole moment of the electron $d_e$ via a two-loop diagram of the Barr-Zee type.
Assuming that the $H$ coupling to the electron is the same as in the Standard Model,
and that there are no other significant contributions to $d_e$, the 
recent upper bound $|d_e/e| < 8.7 \times 10^{-29}$~cm~\cite{Baron:2013eja} 
can be used to set the indirect constraint $|{\tilde \kappa}_t| < 0.01$.
However, we note that there is no experimental information on the electron-$H$ coupling,
that no direct information on this couplings is likely to become available in the foreseeable
future, and that there could in principle be other contributions to $d_e$ that might cancel the two-loop
top contribution, e.g., in supersymmetric models. We therefore seek bounds on $\kappa_t$ and 
${\tilde \kappa}_t$ that are less model-dependent.

As already commented in the Introduction, the data from ATLAS and CMS on $H$
production at the LHC with $E_{CM} =7$ and 8~TeV provide indirect bounds on
the coupling parameters $\kappa_t$ and ${\tilde \kappa}_t$ via the constraints they
impose on the $H$-gluon-gluon and $H$-$\gamma$-$\gamma$ couplings, which
have also been explored in~\cite{Brod:2013cka, Nishiwaki:2013cma}. The interpretation of the $H$-gluon-gluon
and $H$-$\gamma$-$\gamma$ constraints is also somewhat model-dependent,
since they are obtained from data on $H$ production and decay into $\gamma \gamma$
final states, and must rely on assumptions about the $H$ couplings to other particles.
In considering these constraints, we assume here that the couplings to other fermions
and bosons are the same as in the Standard Model, i.e., $\kappa_f = 1$ and 
${\tilde \kappa}_f = 0$ for $f \ne t$, and $\kappa_W = \kappa_Z = 1$.
This assumption is purely phenomenological but motivated by the following reasons.
There are several processes which can constrain
$\kappa_W$ and $\kappa_Z$ independently from $\kappa_t$ and $\tilde \kappa_t$~\cite{Klute:2013cx}
at the time when the luminosity required in this study is accumulated.
The effect of $\kappa_f$ and $\tilde \kappa_f$ ($f \neq t$) is almost negligible unless $\kappa_f \gg 1$
because of the suppression proportional to the Yukawa couplings of the light fermions.
Moreover such a possibility will be ruled out for the bottom and tau by the 
relatively precise $H \to \bar b b$ and $H \to \tau \tau$ measurements available at the time.

Under these assumptions, the available ATLAS and CMS data on $H$ production
and decay were analyzed in~\cite{Ellis:2013lra} and constraints on the $H$-gluon-gluon
and $H$-$\gamma$-$\gamma$ couplings were derived, taking into account the
correlations imposed by the measurements: see the left panel of Fig.~4 of~\cite{Ellis:2013lra}.
The ratios
\begin{equation}
\mu_{gg} \; \equiv \; \frac{\sigma(gg \to H)}{\sigma(gg \to H)_{SM}}, \; \;
\mu_{\gamma \gamma} \; \equiv \; \frac{\Gamma(H \to \gamma \gamma)}{\Gamma(H \to \gamma \gamma)_{SM}}
\label{mus}
\end{equation}
are represented there by $c_g^2$ and $c_\gamma^2$, respectively. Including the
contribution to the $Hgg$ loop amplitude of the $b$ quark and the
contribution to the $H\gamma\gamma$ loop amplitude of the $b$ quark, $\tau$ lepton
and $W$ bosons, following~\cite{Brod:2013cka} one has in the notation of~\cite{Ellis:2013lra}
\begin{eqnarray}
c_g^2 \; = \; \mu_{gg} & \simeq & \kappa_t^2 + 2.6 {\tilde \kappa}_t^2 + 0.11 \kappa_t (\kappa_t - 1) \, , \nonumber \\
c_\gamma^2 \; = \; \mu_{\gamma \gamma} & \simeq & (1.28 - 0.28 \kappa_t)^2 + (0.43 {\tilde \kappa}_t )^2 \, .
\label{cs}
\end{eqnarray}
The left panel of Fig.~4 of~\cite{Ellis:2013lra} displays regions in the $(c_\gamma, c_g)$
plane that are allowed by the LHC data at the 68, 95 and 99\% CL. There we see explicitly
the anticorrelation between $c_g$ and $c_\gamma$ due to the fact that one
may, to some extent, compensate for a possible enhancement in the LHC $H \to \gamma \gamma$
signal~\footnote{We recall that this possibility is suggested by the ATLAS data, but not
by the CMS data, so that the Standard Model value of the $H \gamma \gamma$ coupling is allowed
at the 68\% CL~\cite{Ellis:2013lra}.} by suppressing $\sigma (gg \to H)$, though this possibility is restricted by the LHC
measurements of the strengths of the other $H$ signatures if one assumes that
$\kappa_f = 1$ and  ${\tilde \kappa}_f = 0$ for $f \ne t$, and $\kappa_W = \kappa_Z = 1$
as done here.

\begin{figure}[hbt]
\begin{center}
\includegraphics[height=10cm]{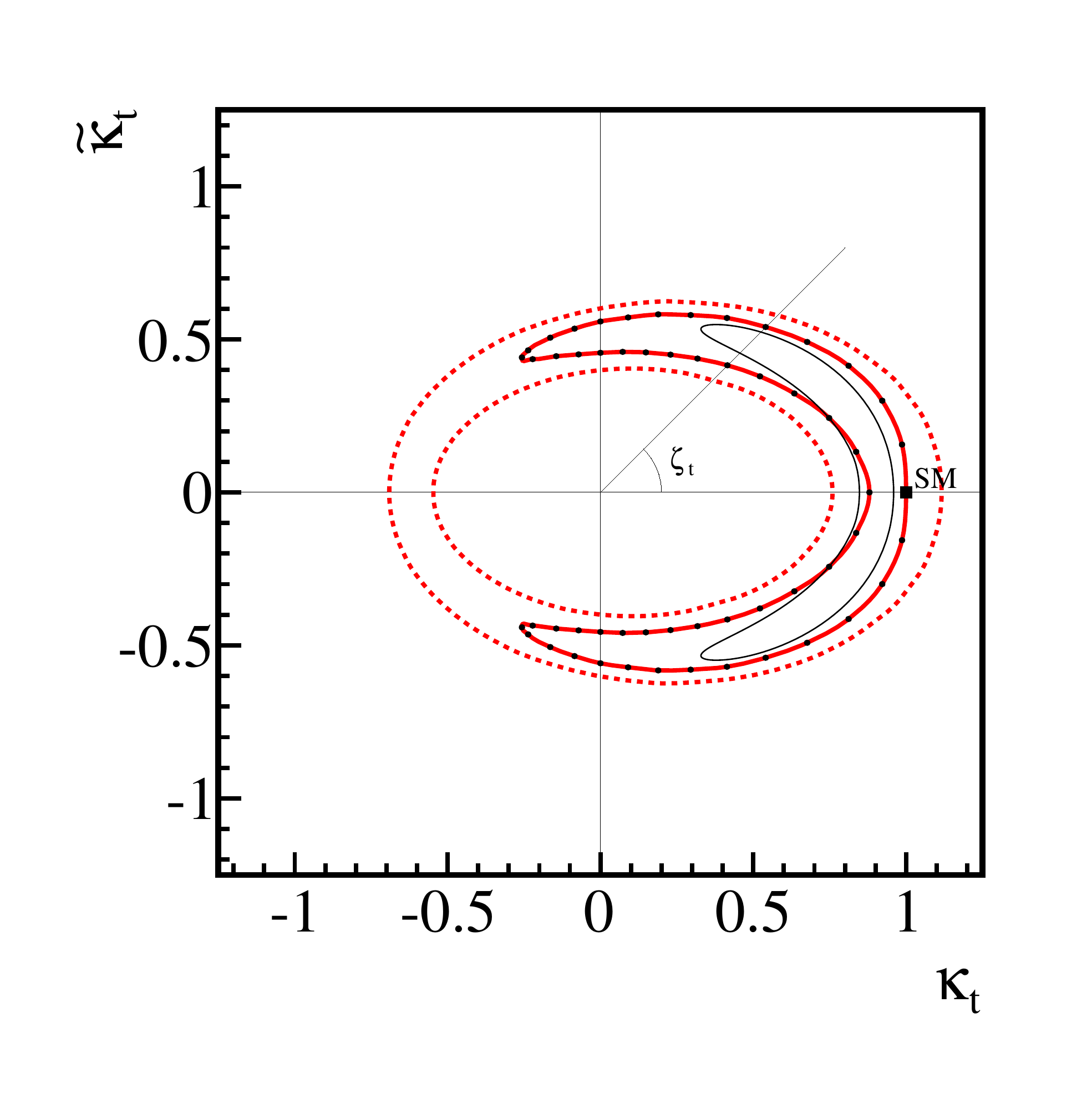}
\end{center}   
\caption{\label{fig:AllowedValues}\it 
The regions of the $(\kappa_t, {\tilde \kappa}_t)$ plane
allowed by the analysis of~\protect\cite{Ellis:2013lra} at the 68 and 95\% CL (solid and dotted red contours,
respectively). Also shown for
comparison is the region discussed in~\protect\cite{Brod:2013cka} (solid black contour).
Black dots represent the simulated model points.
}
\end{figure}

We display in Fig.~\ref{fig:AllowedValues} the regions of the $(\kappa_t, {\tilde \kappa}_t)$
plane that are allowed at the 68, and 95\% CL according to the analysis of~\cite{Ellis:2013lra}. 
At the 68\% CL, the allowed region is a crescent with apex close to the Standard Model point
$(\kappa_t, {\tilde \kappa}_t) = (1, 0)$, bounded by the solid red contour, 
whereas at the 95\% CL a complete annulus is allowed, bounded by the dotted red contour. 
For convenience we define the CP violation phase in the $\bar{t}t H$ coupling by
\begin{equation}
\zeta_t \; \equiv \; \arctan \Big( \frac{\tilde \kappa_t}{\kappa_t} \Big) \,.
\end{equation}
For comparison, we also display the (smaller) crescent discussed in~\cite{Brod:2013cka, Nishiwaki:2013cma}, bounded by the solid black contour. 
As already mentioned, if one assumes the Standard Model value of the electron-$H$ coupling and there are no other
important contributions to the EDM of the electron $d_e$, the experimental upper limit on its
value imposes $|{\tilde \kappa}_t| < 0.01$. Here we consider the capability of future LHC
measurements to constrain $\kappa_t$ and ${\tilde \kappa}_t$ directly, considering for
illustration the full crescent allowed by the analysis of~\cite{Ellis:2013lra} at the 68\% CL.

\section{Total Cross Sections}

We have simulated the production of ${\bar t} t H$, $t H$ and ${\bar t} H$ final states at
the LHC in leading order~\footnote{Evaluating NLO corrections lies beyond the scope of
this work, but we do not expect them to alter qualitatively the
results found here.} using {\tt MadGraph}~\cite{Alwall:2011uj} 
All the results presented below are for a centre-of-mass energy of 14~TeV.
We consider first the effects on the total cross sections for $H$ production in association
with ${\bar t}t$, single $t$ and single ${\bar t}$, taking into account the present LHC constraints
discussed in the previous Section, and discuss other possible LHC measurements in subsequent Sections.

\subsection{Cross Sections for ${\bar t}tH$ Production}

\begin{figure}[t!]
\begin{center}
\includegraphics[height=7cm]{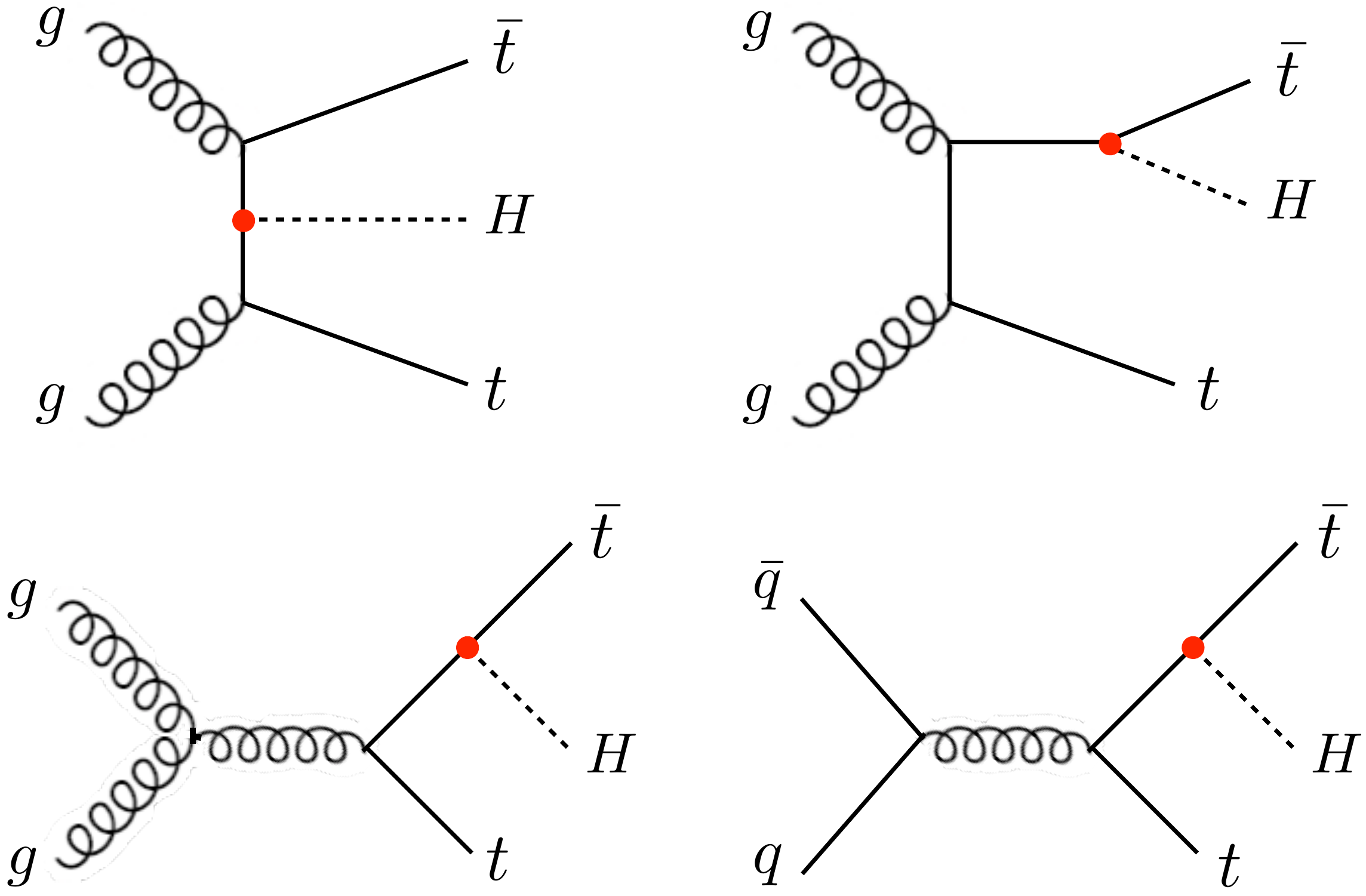}\\
(a) \\
\end{center}
\vspace{0.5cm}
\begin{center}
\includegraphics[height=3.5cm]{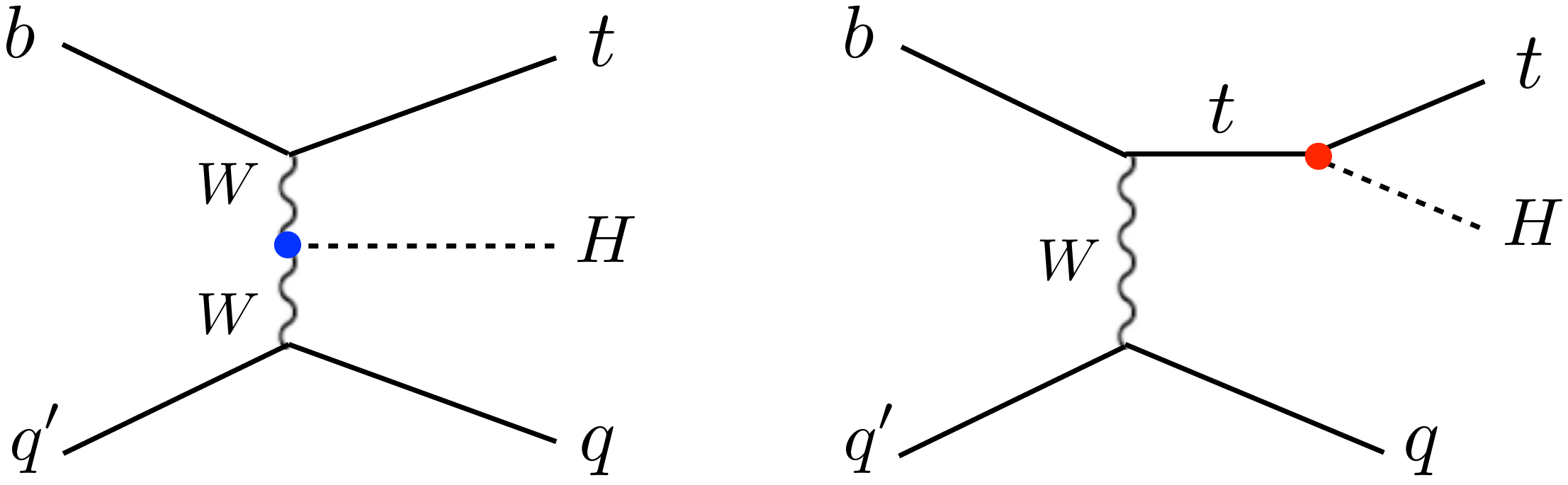} \\
(b) \\
\end{center}   
\caption{\label{fig:diagrams}\it
Leading diagrams
contributing to ${\bar t}tH$ production at the LHC (upper panel) and
to single $t$ or ${\bar t}$ production (lower panel).
The red and blue dots correspond to the $\bar t t H$ and $WWH$ couplings, respectively. 
}
\end{figure}

\begin{figure}[hbt]
\begin{center}
\includegraphics[height=8cm]{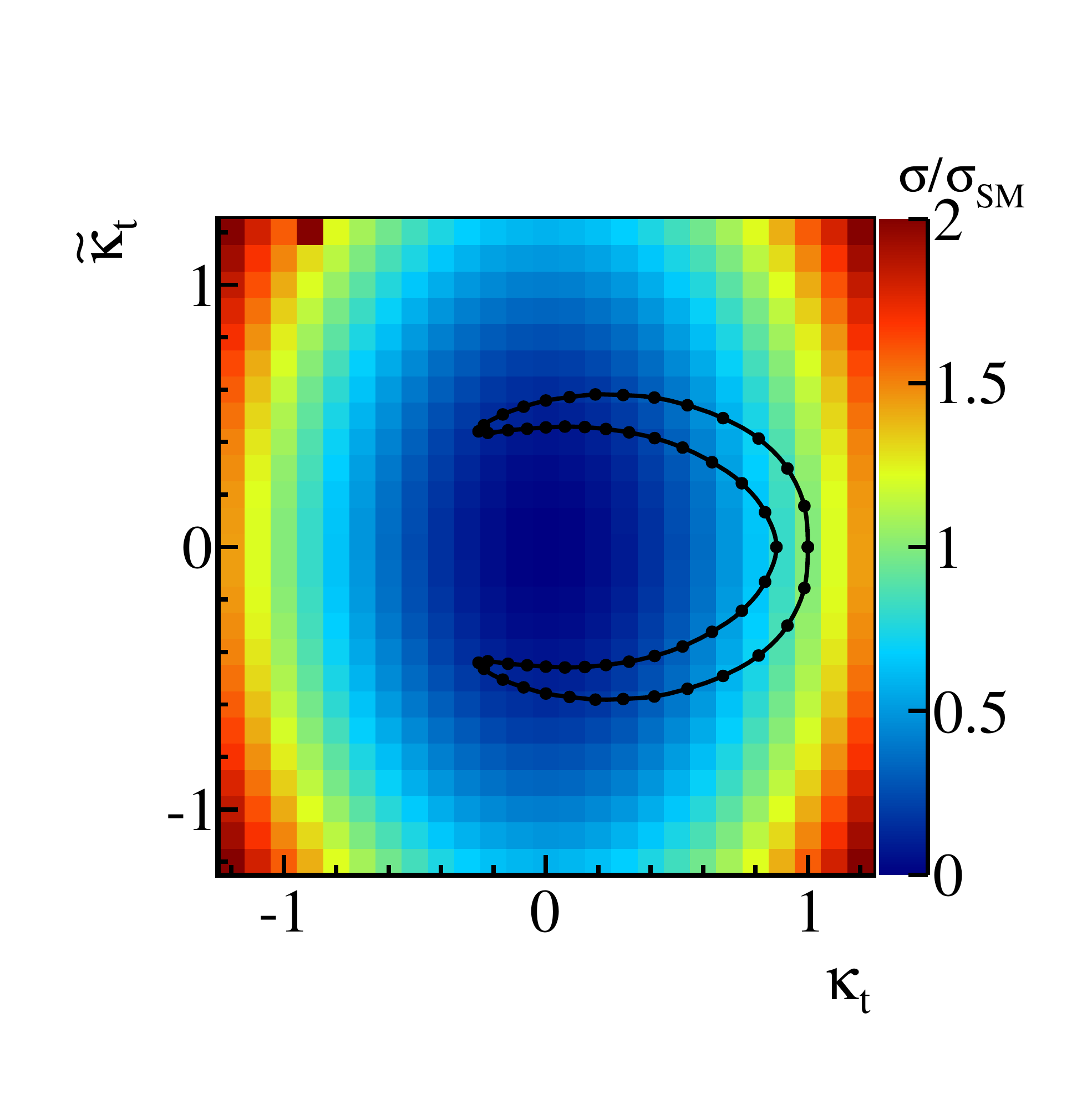}
\includegraphics[height=8cm]{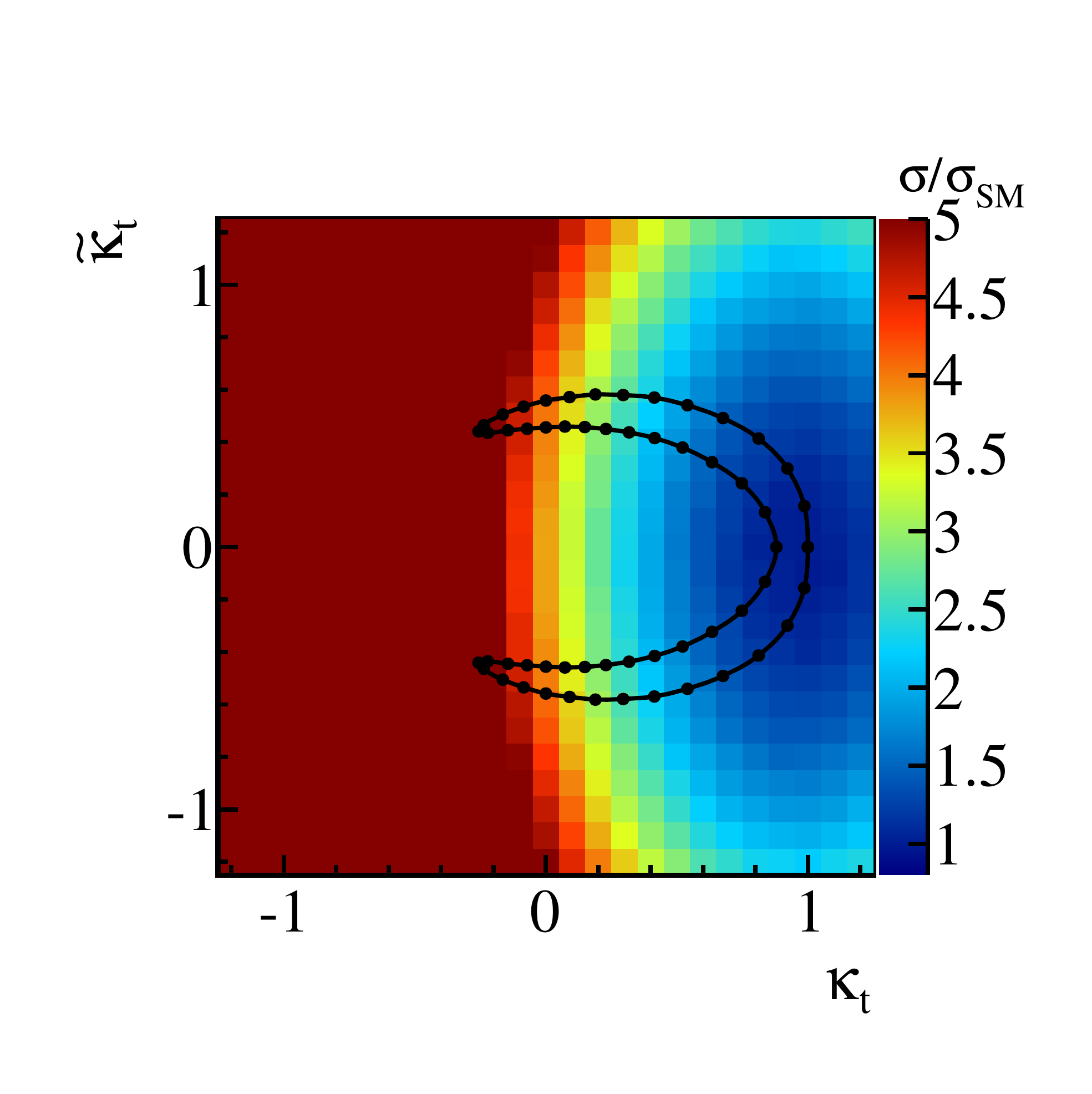} \\
\end{center}   
\caption{\label{fig:Ratios}\it
The ratios of $\sigma({\bar t}tH)$ to the Standard Model value (left panel) and
of $\sigma(tH)$ to the Standard Model value (right panel)
are shown using the indicated colour codes. Also shown is the crescent-shaped region in 
Fig.~\protect\ref{fig:AllowedValues} that is allowed by present data at the 68\% CL.
}
\end{figure}

The leading tree-level diagrams for ${\bar t}tH$ production at the LHC are displayed in the upper
panel of Fig.~\ref{fig:diagrams}, and the left panel of Fig.~\ref{fig:Ratios} displays
the corresponding values of $\sigma({\bar t}tH)$ at the LHC at 14~TeV, using colour-coding
to represent the ratio to the Standard Model cross section.
The contributions of the ${\bar t}t H$ and ${\bar t} \gamma_5 t H$ couplings to $\sigma({\bar t}tH)$
do not interfere, so the iso-$\sigma$ contours are ellipses in the $(\kappa_t, {\tilde \kappa}_t)$
plane. For equal values of $\kappa_t$ and ${\tilde \kappa}_t$, the latter yields a smaller
cross section, with the result that the major axes of these ellipses are aligned with the 
${\tilde \kappa}_t$ axis, as seen in the left panel of Fig.~\ref{fig:Ratios}. Also shown there is
the crescent-shaped region allowed by the present LHC data at the 68\% CL. It is clear that
in this region $\sigma({\bar t}tH)$ is in general {\it smaller} than in the Standard Model, as we
discuss in more detail later.

\begin{figure}[hbt]
\begin{center}
\includegraphics[height=8cm]{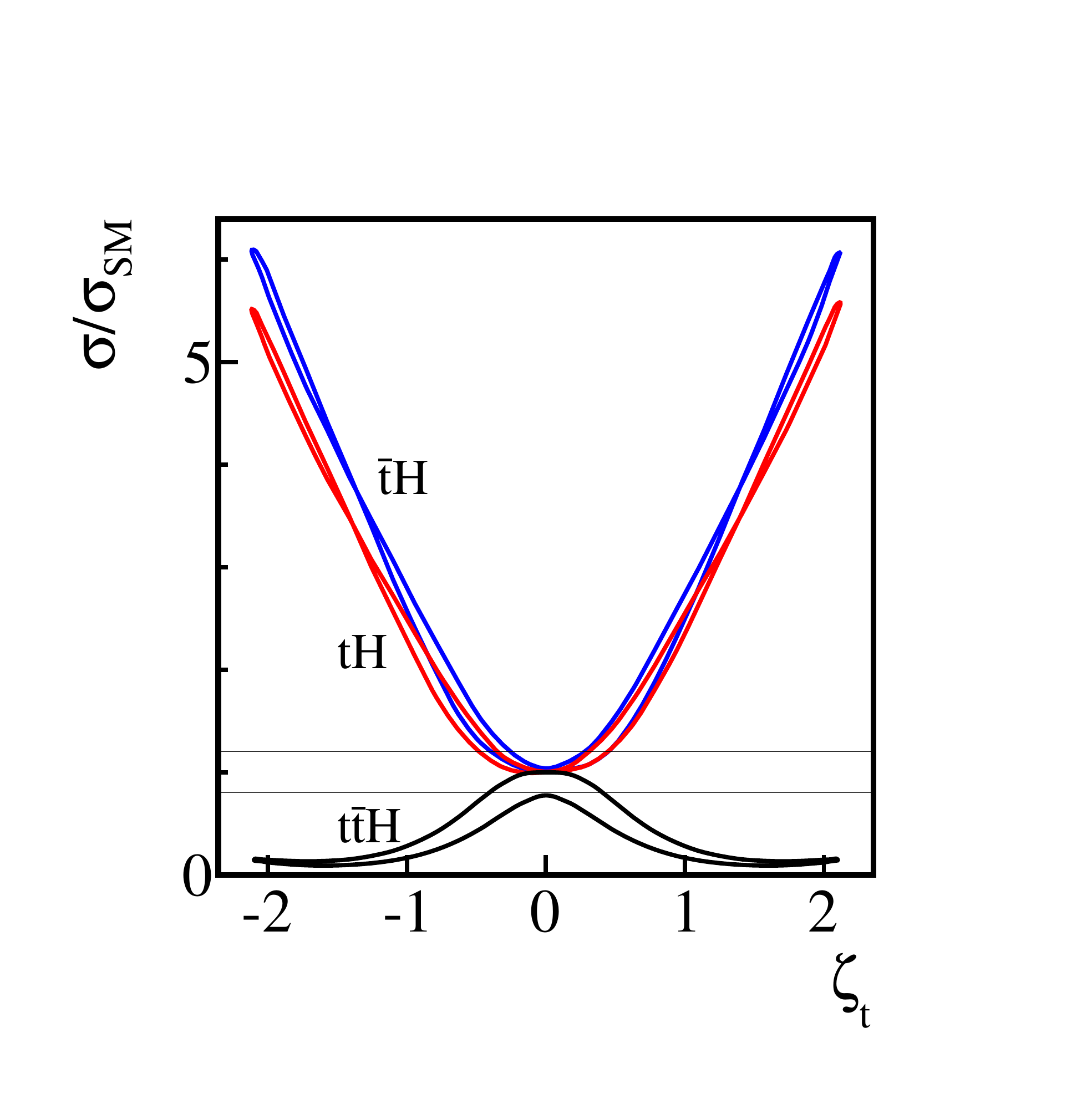}
\includegraphics[height=8cm]{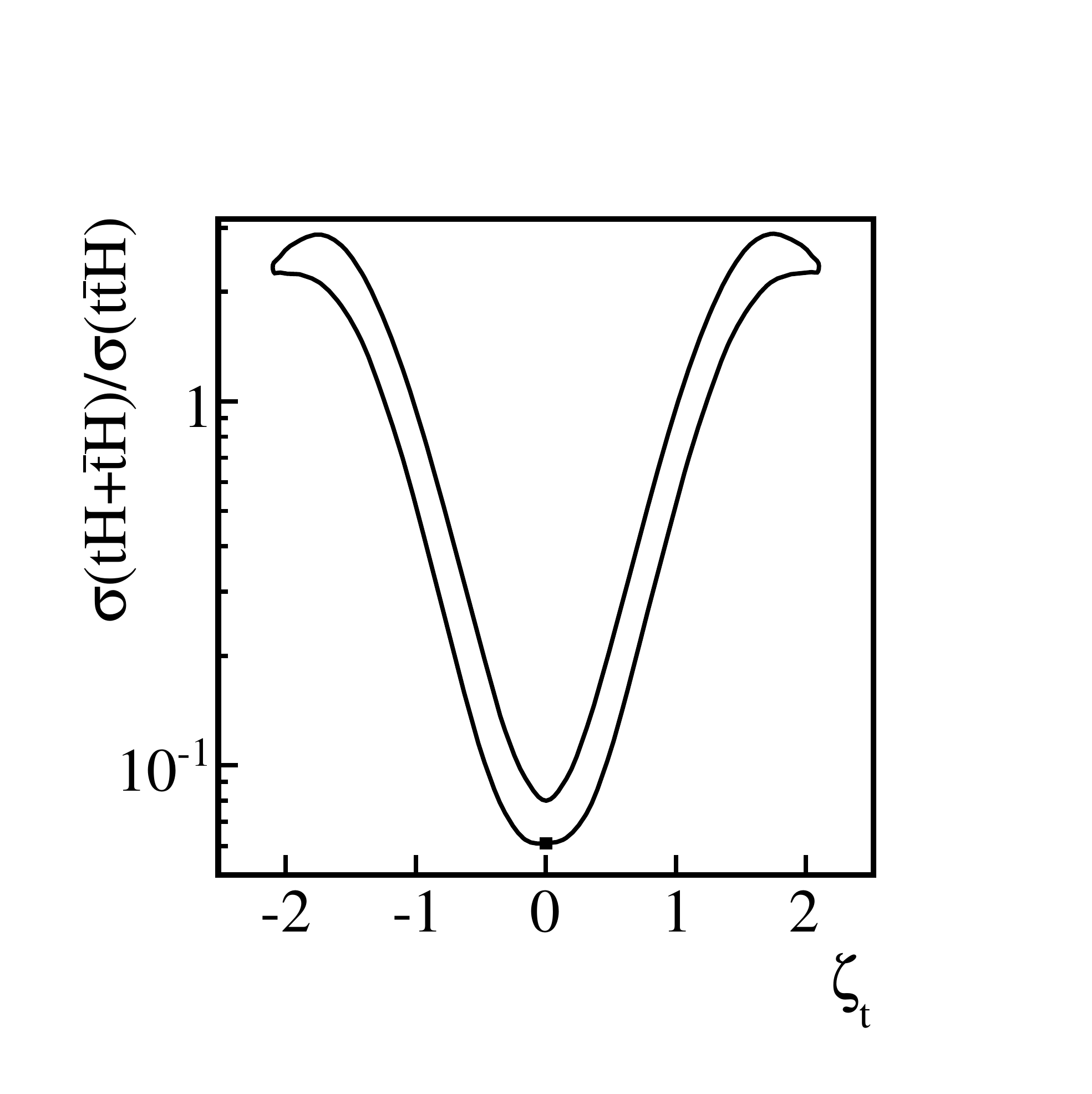} \\
\end{center}   
\caption{\label{fig:RatioRanges}\it
Left panel: The ratios $\sigma({\bar t}tH)/\sigma({\bar t}tH)_{SM}$ (black lines), $\sigma(tH)/\sigma(tH)_{SM}$ (red lines)
and $\sigma({\bar t}H)/\sigma({\bar t}H)_{SM}$ (blue lines) as functions of ${\rm arc} \tan ({\tilde \kappa}_t/ \kappa_t)$.
Right panel: The ratio $\sigma(tH) + \sigma({\bar t}H)/\sigma({\bar t}tH)$ as a function of
${\rm arc} \tan ({\tilde \kappa}_t/ \kappa_t)$. In both panels, we display the values of the ratios
along both the inner and outer boundaries of the crescent-shaped region in 
Fig.~\protect\ref{fig:AllowedValues} that is allowed by present data at the 68\% CL.
The horizontal lines in the left panel correspond to a measurement of the cross section for ${\bar t} t H$
at the Standard Model level with an accuracy of 20\%.
}
\end{figure}

The left panel of Fig.~\ref{fig:RatioRanges} displays the variation of
the ratio $\sigma({\bar t}tH)/\sigma({\bar t}tH)_{SM}$ along the boundary of
the 68\% CL crescent displayed in Fig.~\ref{fig:AllowedValues}. 
The horizontal axis is the CP violation phase, $\zeta_t$, which parametrizes the boundary, and the upper and lower lines
correspond to the outer and inner boundaries of the crescent, respectively.
We see that an LHC measurement of $\sigma({\bar t}tH)$ could in principle put an
interesting constraint on $\zeta_t$. For example, a
measurement at the Standard Model level with an accuracy of 20\%, indicated by the horizontal lines in the
left panel of Fig.~\ref{fig:RatioRanges}, would determine
$\zeta_t \sim 0 \pm 30^o$.

\subsection{Cross Sections for $tH$ and ${\bar t}H$ Production}

We now discuss the total cross sections for the associated production of $H$ with a single $t$ or ${\bar t}$
and a light-quark jet via the tree-level diagrams are shown in the lower panel of Fig.~\ref{fig:diagrams}.
We neglect $s$-channel ${\bar q} q \to t H {\bar b}$ and ${\bar t} H b$ production, 
since their cross sections are an order of magnitude smaller than the processes with light quarks~\cite{Maltoni:2001hu}.
In Fig.~\ref{fig:diagrams}, we note that $H$ may be radiated
from either a final-state $t$ quark or an intermediate virtual $W$ boson. It
has been noticed previously that the interference between these diagrams is 
sensitive to the relative magnitude and sign of the scalar ${\bar t } t H$ and
$WWH$ couplings, with the result that $\sigma(tH)$ and $\sigma({\bar t}H)$
are minimized around the Standard Model value $\kappa_t = 1$~\cite{Maltoni:2001hu}.\footnote{ 
Disturbing the $\bar t t H$ coupling modifies the UV behaviour of the theory and
may lead to a violation of the perturbative unitarity at some scale $\Lambda_{\rm UV}$.
It has been shown in \cite{Farina:2012xp} that this effect is most pronounced at $\kappa_t = -1$ but $\Lambda_{\rm UV} \ga 9$~TeV
even in that case.    
This implies that the perturbative calculation used in our paper is still reliable.
}
As in the case of $\sigma({\bar t}tH)$, iso-$\sigma$ contours for $t H$ and ${\bar t}H$
production are also ellipses whose major axes are aligned with the 
${\tilde \kappa}_t$ axis, as we see in
the right panel of Fig.~\ref{fig:Ratios}, where colour-coding is used to 
represent the ratio to the Standard Model cross section. As a consequence, $\sigma(tH)$ and 
$\sigma({\bar t}H)$ {\it increase} along the 68\% CL crescent as $\kappa_t$
decreases and ${\tilde \kappa}_t$ increases in magnitude.

This effect is also seen clearly in the left panel of Fig.~\ref{fig:RatioRanges}, where
we see that $\sigma(tH)$ and $\sigma({\bar t}H)$ reach more than 3 times
the Standard Model values when $\zeta_t > 60^o$. A measurement at the Standard Model level with a factor of two
uncertainty would determine $\zeta_t \sim 0 \pm 45^o$.
As seen in the right panel of Fig.~\ref{fig:RatioRanges}, the combination of
the decrease in $\sigma({\bar t}tH)$ and the increases in $\sigma(tH)$ and 
$\sigma({\bar t}H)$ along the crescent imply that the ratio 
$\sigma(tH + {\bar t}H)/\sigma({\bar t}tH)$ increases by a factor of more than 20
along the crescent, compared to its value in the Standard Model, $\sim 0.06$.

\section{Mass Distributions}

We now examine the information that can be obtained from measurements
of the invariant masses of combinations of the final-state $t$, ${\bar t}$ and $H$
particles. In the case of the ${\bar t}tH$ final state, there are three distinct
combinations that can be measured: the total invariant mass $M_{{\bar t}tH}$,
the $tH$ (or ${\bar t}H$) invariant mass $M_{tH}$ (or $M_{{\bar t}H}$), and the
${\bar t}t$ invariant mass $M_{{\bar t}t}$. In the case of single $t$ or ${\bar t}$
production, there is also a forward jet $j$ corresponding to the quark from
which the virtual $W$ was emitted, as seen in the lower panel of Fig.~\ref{fig:diagrams}.
Hence there are again three final-state particles $t$ (or ${\bar t}$), $H$ and $j$,
and therefore four measurable invariant masses in this case: the total invariant
mass $M_{tHj}$ (or $M_{{\bar t}Hj}$) and the two-particle invariant masses
$M_{tH}$ (or $M_{{\bar t}H}$), $M_{tj}$ (or $M_{{\bar t}j}$), and $M_{Hj}$.
In the following we present some invariant mass distributions for the ${\bar t}tH$
and $tHj$ (or ${\bar t}Hj$) final states, starting with the total invariant mass distributions.
All the distributions shown below are {\it idealized}, as they 
do not include the effects of parton showering, object reconstruction, detector resolution, etc..
We also do not consider the background contamination and the realistic selection cuts which will be applied in experiments.\footnote{
The background contamination is known to be a serious problem for the $\bar t t H$ process.
In addition to improving the techniques to suppress the the background, e.g using jet substructure techniques~\cite{Plehn:2009rk},
a precise estimation of the background shapes would be necessary to reduce the systematic uncertainties.  
}
These effects could alter the shape of distributions, but 
the study of such effects lies beyond this exploratory work.

\subsection{Total Invariant Mass Distributions}

\begin{figure}
\begin{center}
\includegraphics[height=5.2cm]{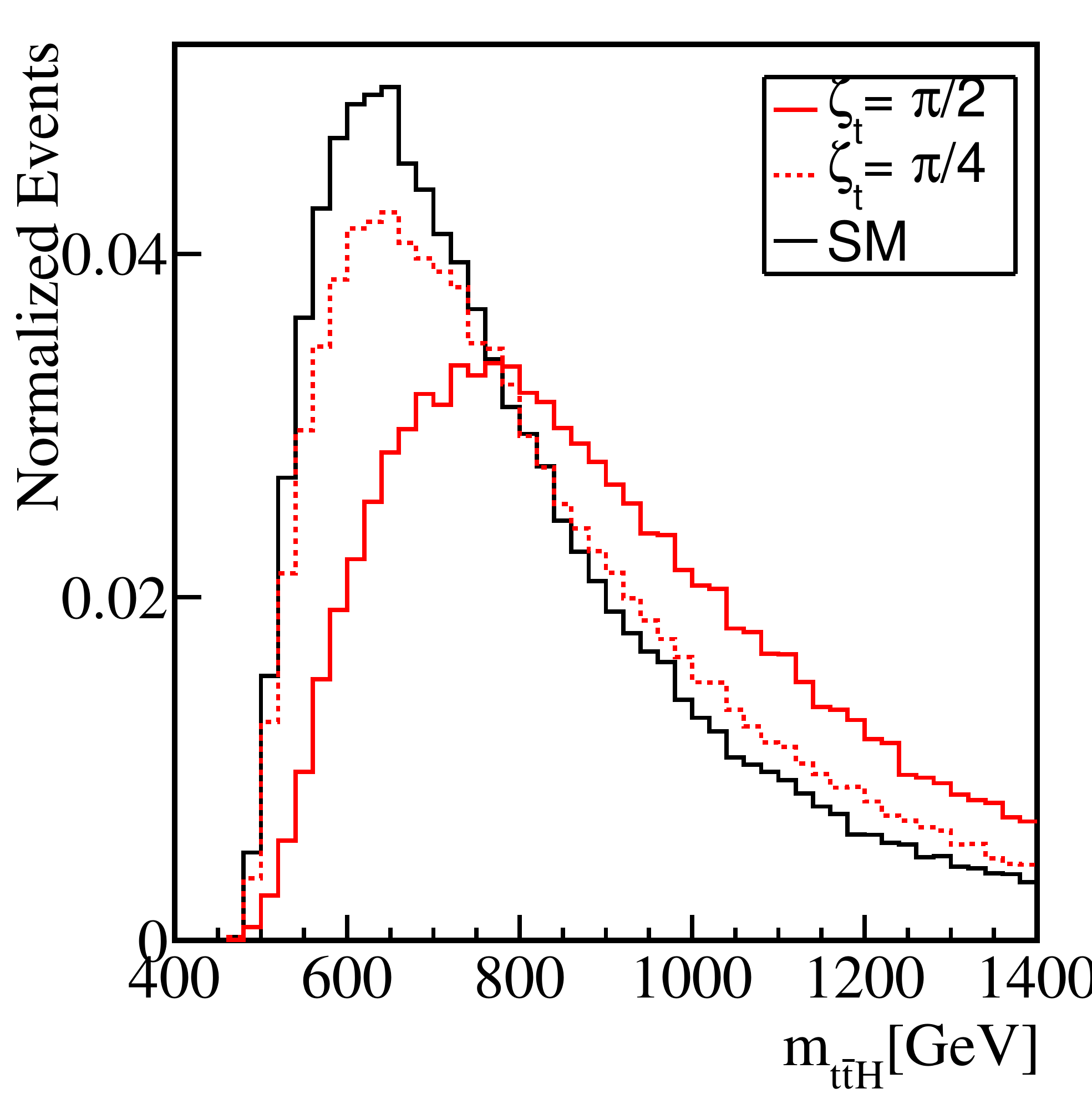}
\includegraphics[height=5.2cm]{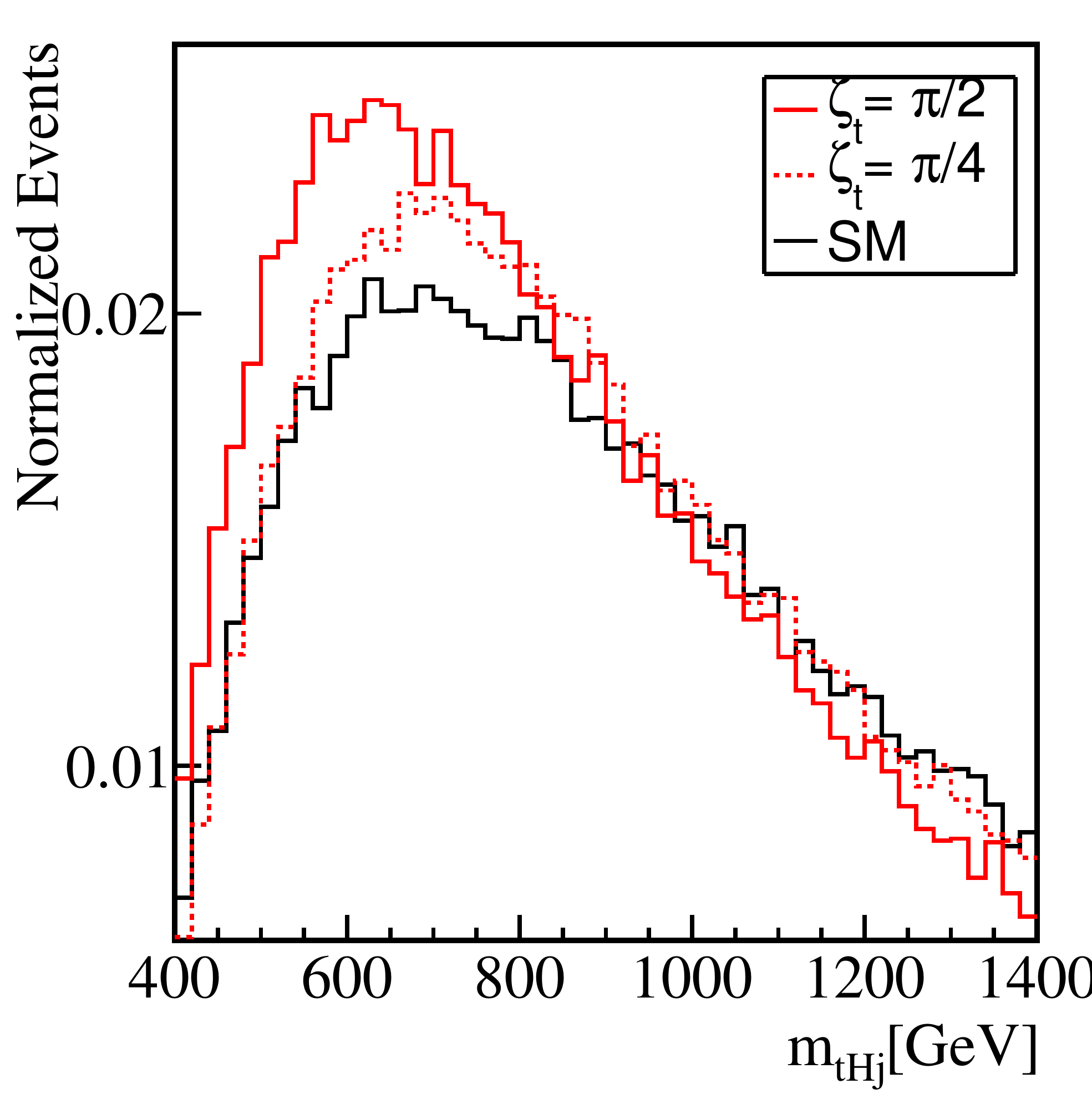} 
\includegraphics[height=5.2cm]{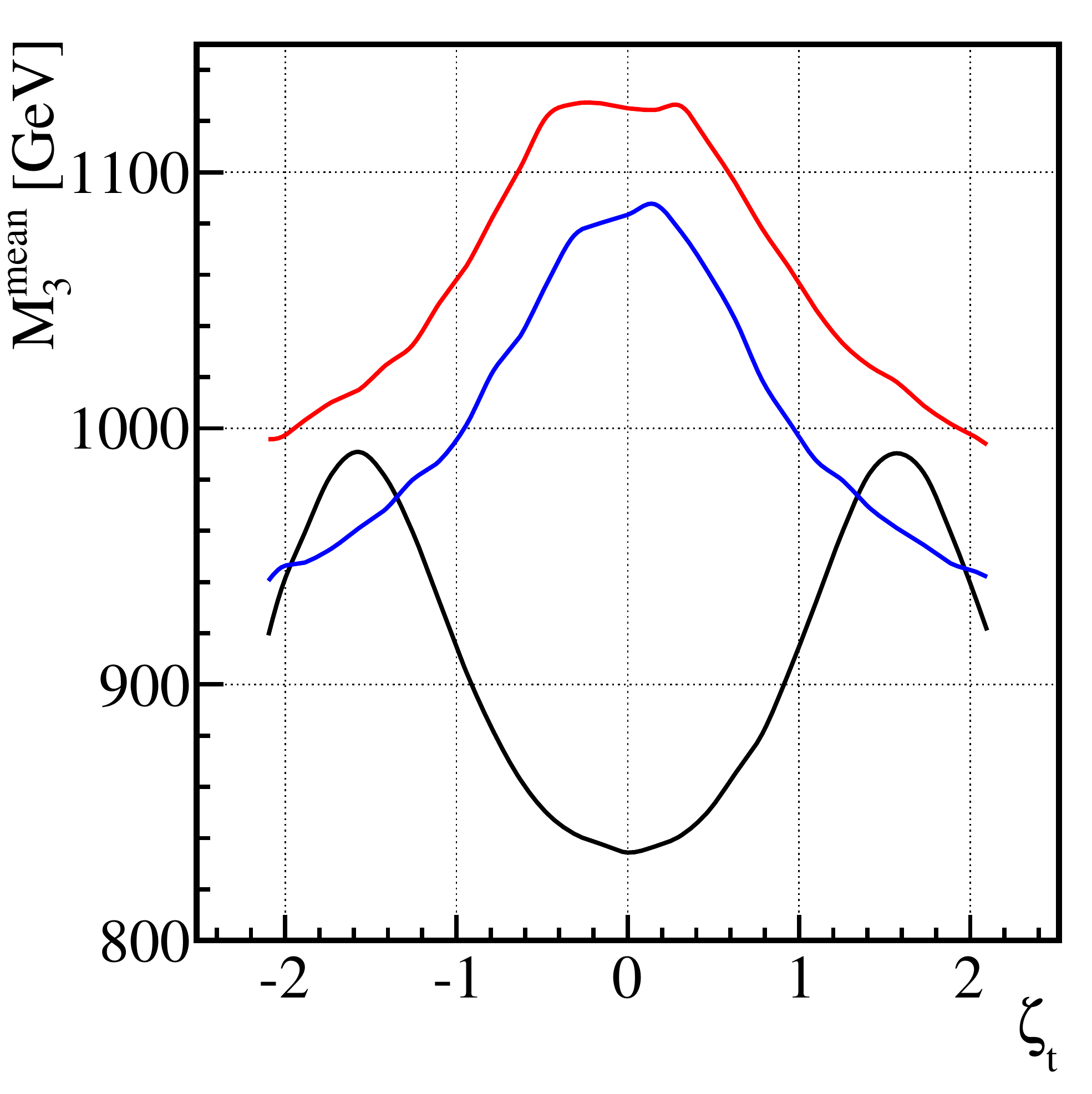}
\end{center}   
\caption{\label{fig:mass}\it
The total invariant mass distributions for the ${\bar t}tH$ final state (left panel)
and the $tHj$ final state (central panel). In each case, we display the distributions
for $\zeta_t = {\rm arc} \tan ({\tilde \kappa}_t/ \kappa_t) = 0$ (in black), $\pm \pi/4$
(in dotted red) and $\pm \pi/2$ (in solid red). The right panel shows the variations with $\zeta_t$ of
$\langle M_{{\bar t}tH} \rangle$ (solid black), $\langle M_{tHj} \rangle$ (solid red) and 
$\langle M_{{\bar t}Hj} \rangle$ (solid blue) along a contour passing trough the middle 
of the 68\% CL. crescent-shape allowed region in Fig.~\ref{fig:AllowedValues}.}
\end{figure}

The left panel of Fig.~\ref{fig:mass} displays the normalized $M_{{\bar t}tH}$ distributions
for $\zeta_t = {\rm arc} \tan ({\tilde \kappa}_t/ \kappa_t) = 0$ (in black), $\pm \pi/4$
(in dotted red) and $\pm \pi/2$ (in solid red). We see that the
$M_{{\bar t}tH}$ distribution that is {\it most} peaked towards small masses
is that for the Standard Model case $\zeta_t = 0$. That for $\zeta_t = \pm \pi/4$ is less peaked,
and that for $\zeta_t = \pm \pi/2$ is substantially broader.

The central panel of Fig.~\ref{fig:mass} displays the $M_{tHj}$ distributions
for $\zeta_t = 0, \pm \pi/4$ and 
$\pm \pi/2$ using the same colour-coding. In this case, we see that the
invariant mass distribution is {\it least} peaked for the Standard Model case $\zeta_t = 0$,
more peaked for $\zeta_t = \pm \pi/4$ and particularly for $\zeta_t = \pm \pi/2$.

The right panel of Fig.~\ref{fig:mass} displays the variations with $\zeta_t$ of
$\langle M_{{\bar t}tH} \rangle$ (solid black), 
$\langle M_{tHj} \rangle$ (solid red) and 
$\langle M_{{\bar t}Hj} \rangle$ (solid blue). 
We see explicitly that $\langle M_{{\bar t}tH} \rangle$ is {\it minimized}
in the Standard Model case, whereas $\langle M_{tHj} \rangle$ and $\langle M_{{\bar t}Hj} \rangle$
are {\it maximized} in this case. These features are correlated with the behaviours of the total cross
sections for these processes as functions of $\zeta_t$. We note that $\langle M_{{\bar t} t H} \rangle$
is {\it maximized} for $|\zeta_t| = \pi/2$: the value for $|\zeta_t| = \pi$ would be the same as in the Standard Model.

\subsection{Two-Particle Invariant Mass Distributions}

More information may be obtained from two-particle invariant mass distributions,
and we start by showing the two-body mass distributions in ${\bar t} t H$ production events.
\begin{figure}
\begin{center}
\includegraphics[height=5.2cm]{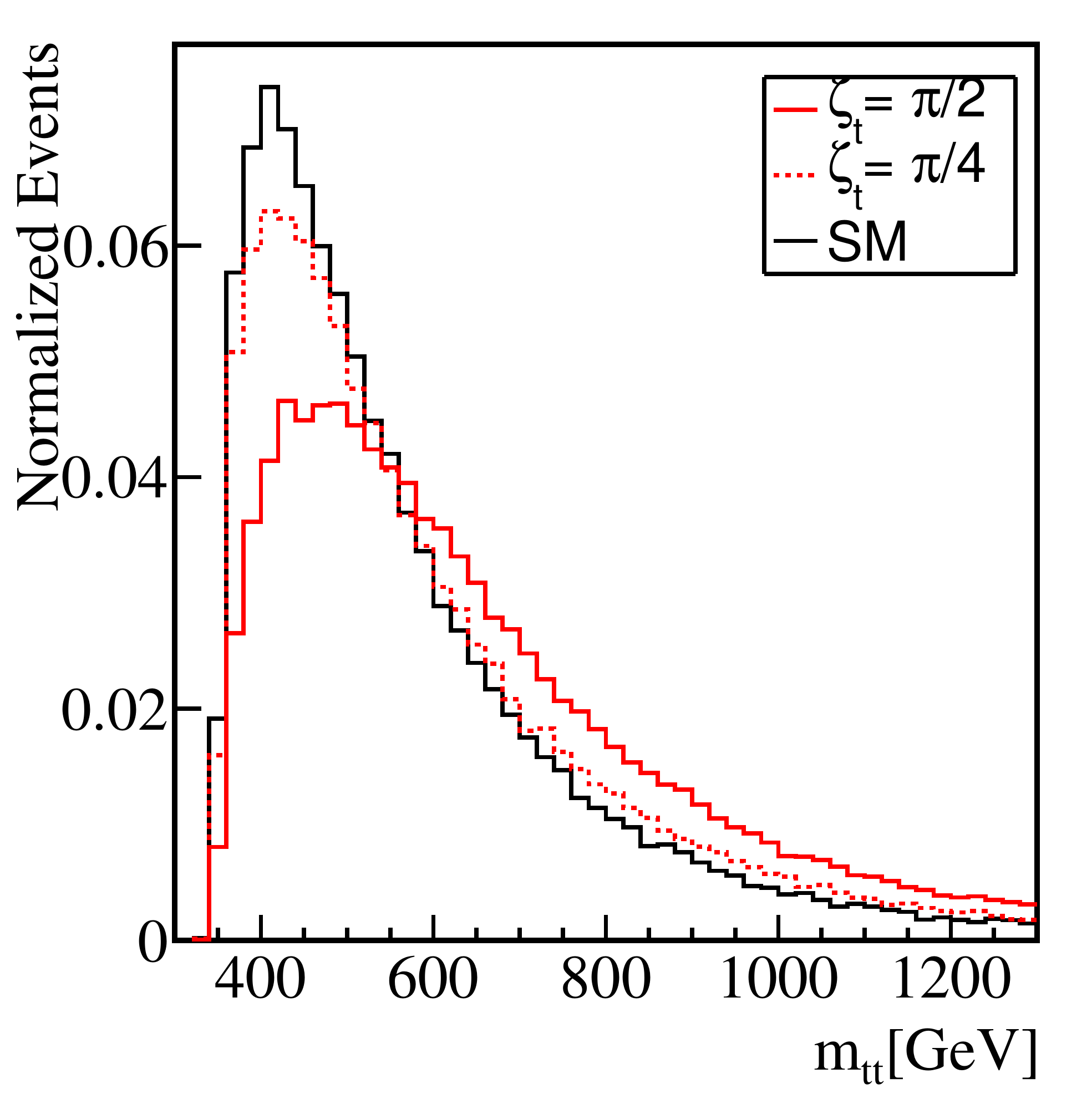}
\includegraphics[height=5.2cm]{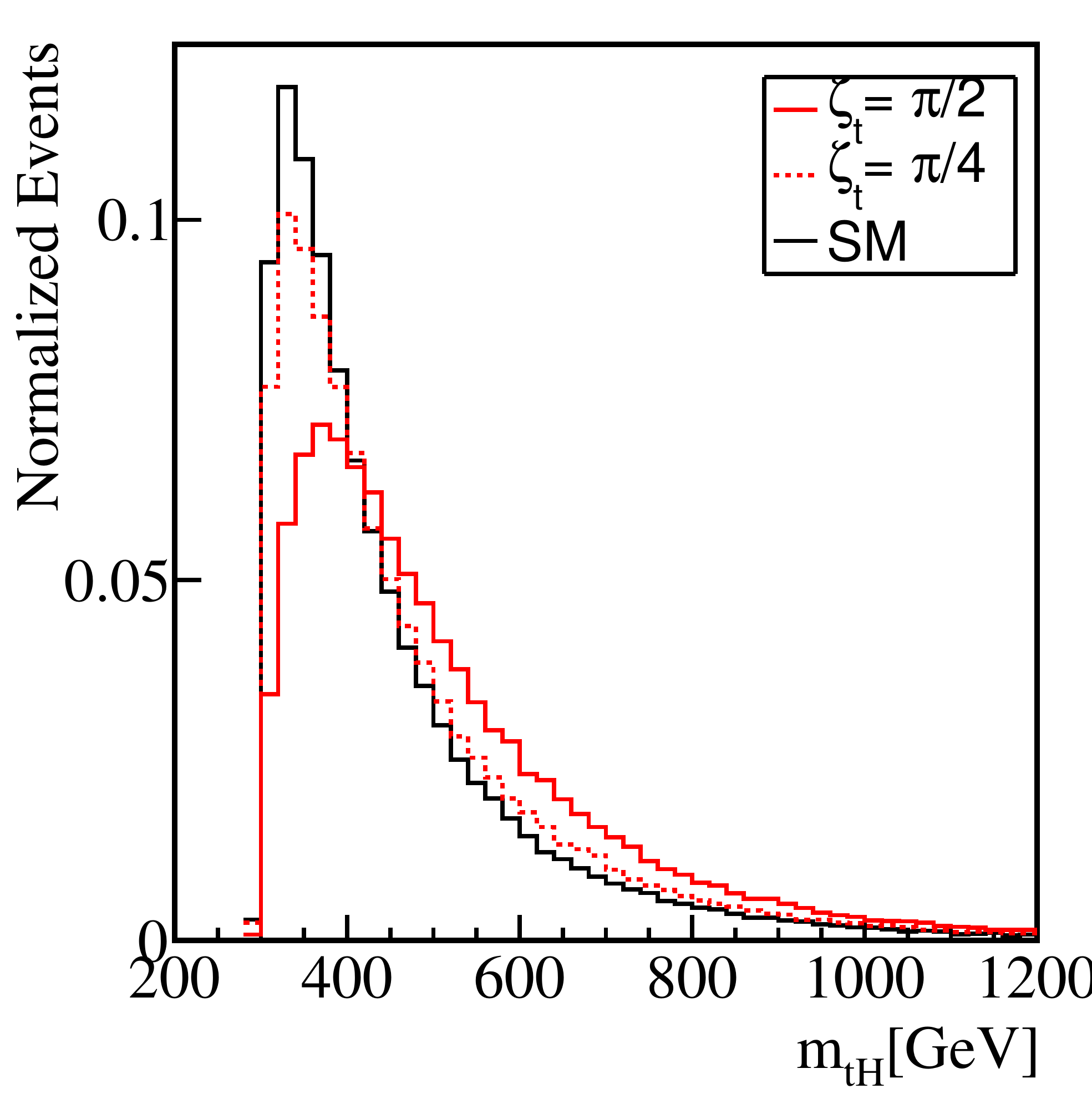}
\includegraphics[height=5.2cm]{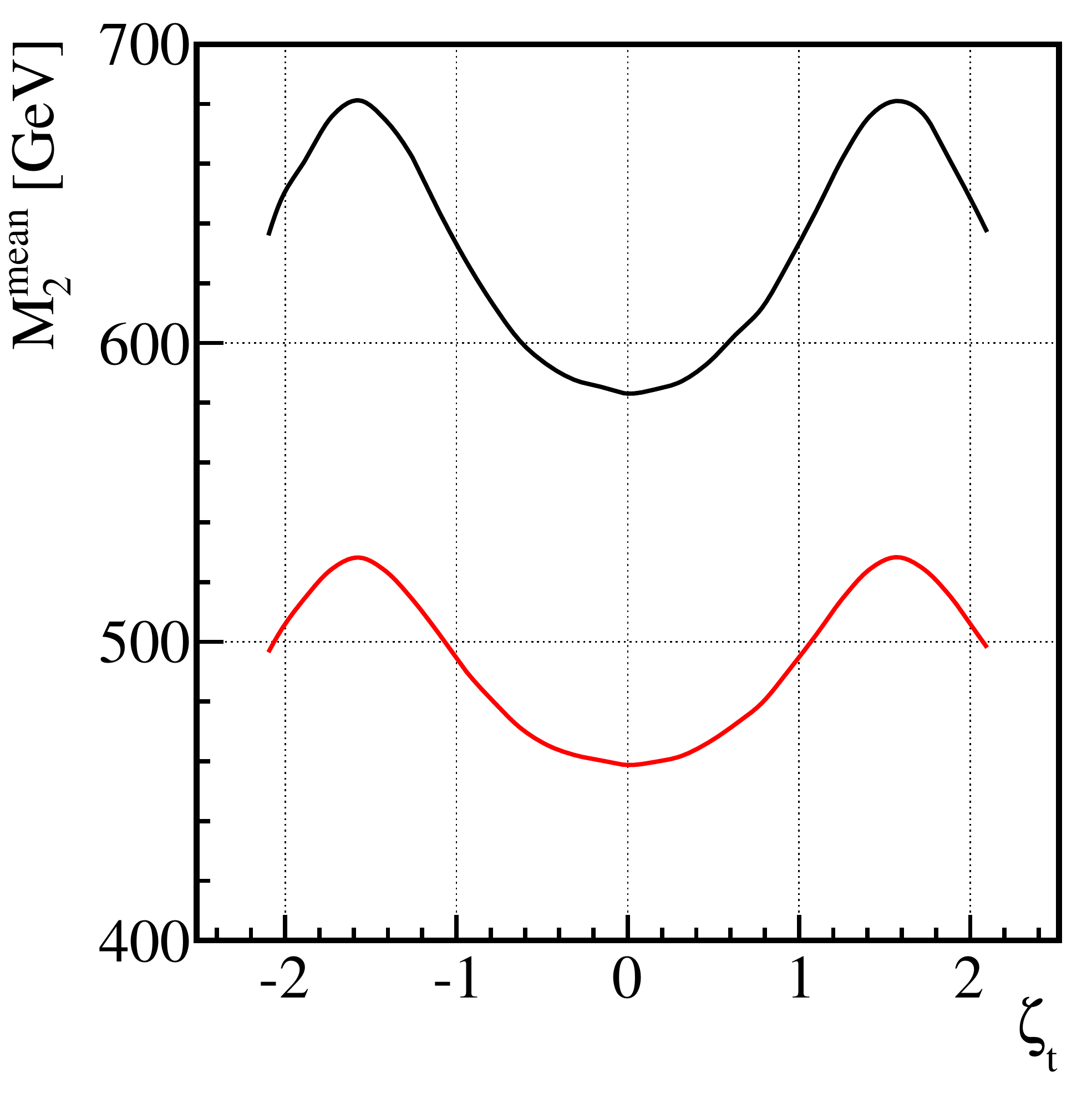}
\end{center}   
\caption{\label{fig:m2_tbarth}\it
The invariant mass distributions of $\bar{t} t$ (left panel) and $t H$ (central panel) in $\bar{t} t H$ production events.
In each case, we display the distributions
for $\zeta_t = {\rm arc} \tan ({\tilde \kappa}_t/ \kappa_t) = 0$ (in black), $\pm \pi/4$
(in dotted red) and $\pm \pi/2$ (in solid red). 
The right panel shows the variations with $\zeta_t$ of $\langle M_{\bar t t} \rangle$ (solid black) 
and $\langle M_{t H} \rangle$ (solid red)
along a contour passing trough the middle of the 68\% CL. crescent-shape allowed region in Fig.~\ref{fig:AllowedValues}.
}
\end{figure}
The left and central panels of Fig.~\ref{fig:m2_tbarth} show the invariant mass distributions of $\bar t t$ and $t H$, 
respectively, with the same colour-coding as in Fig.~\ref{fig:mass}.
The peaks of the distributions are lowest for the SM and highest for $\zeta_t = \pm \pi/2$ in both the $\bar t t$ and $t H$ cases.  
The right panel of Fig.~\ref{fig:m2_tbarth} shows
the variation with $\zeta_t$ of $\langle M_{\bar t t} \rangle$ (solid black) and $\langle M_{t H} \rangle$ (solid red)
along a contour passing trough the middle of the crescent-shape allowed region in Fig.~\ref{fig:AllowedValues}.
The means of the two-particle invariant mass distributions take their
{\it lowest} values in the Standard Model case and their {\it maximum} values for $\zeta_t = \pm \pi/2$
in both the $\bar t t$ and $t H$ cases, as observed in the total invariant mass distribution. 
The difference between $\langle M_{\bar t t} \rangle$ and $\langle M_{t H} \rangle$ is more than 100~GeV, 
despite the difference between $m_t$ and $m_H$ being less than $50$~GeV,  
and is almost independent of $\zeta_t$.
We do not show the $\bar t H$ invariant mass distribution as it is identical to that for $ t H$.

We now turn to the two-particle invariant mass distributions in the $tHj$ and $\bar t Hj$ production events.
\begin{figure}
\begin{center}
\includegraphics[height=5.2cm]{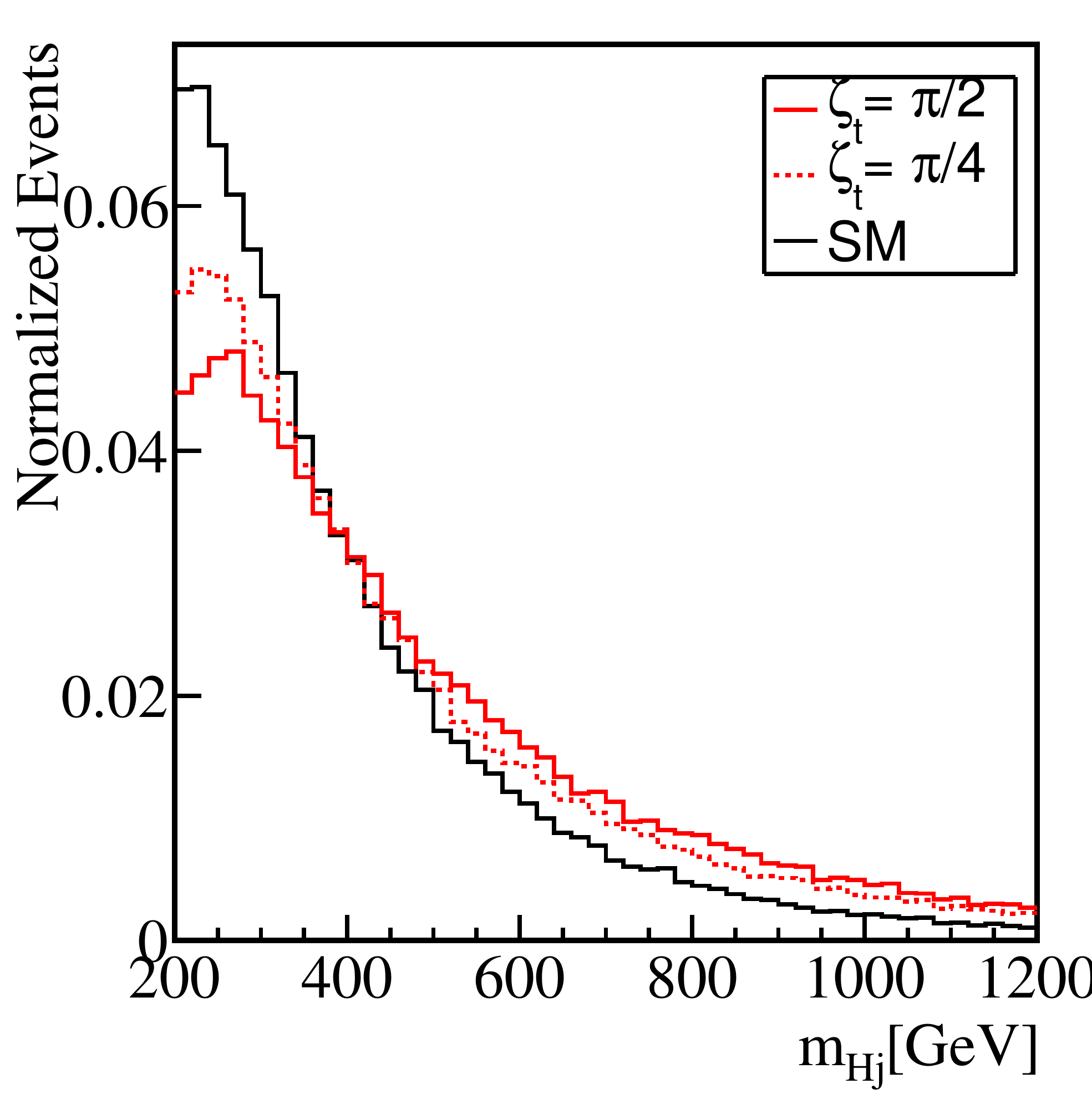}
\includegraphics[height=5.2cm]{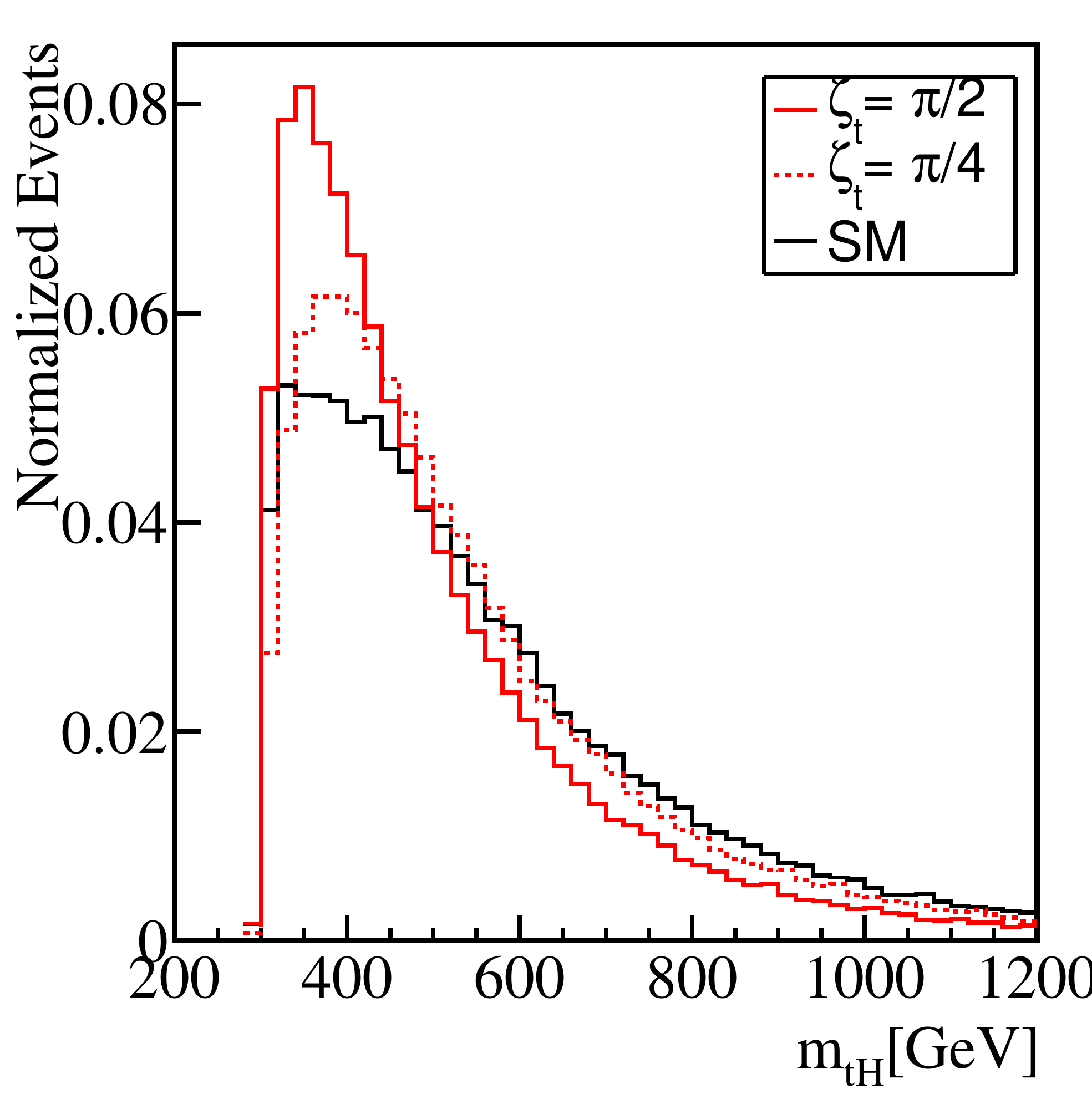}
\includegraphics[height=5.2cm]{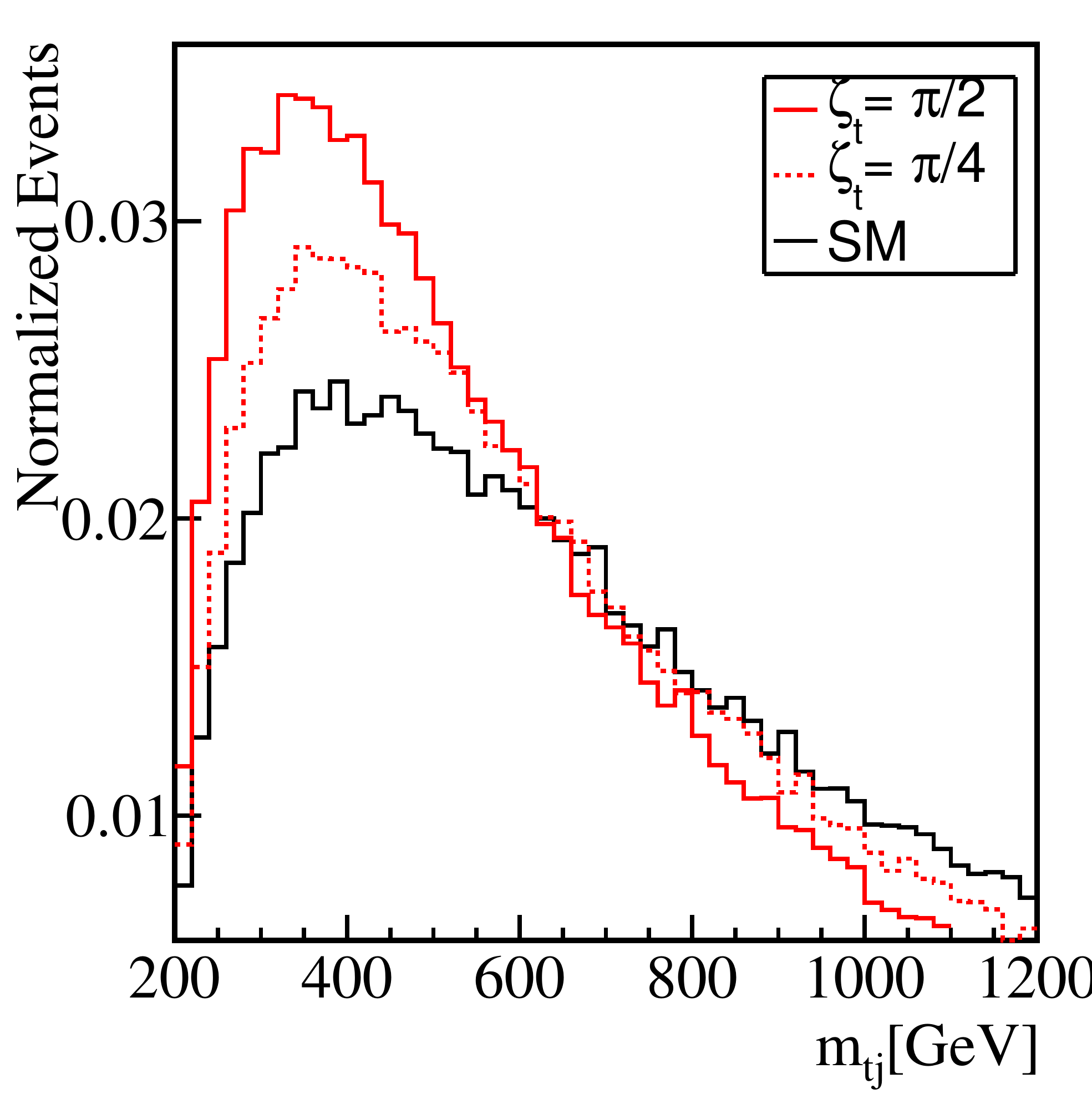}
\end{center}   
\caption{\label{fig:m_tbarth}\it
The invariant mass distributions of $Hj$ (left panel), $t H$ (central panel) and $tj$ (right panel) in $ t H j$ production events.
The black solid, red dashed and red solid histograms correspond to $|\zeta_t| = |\arctan \tilde \kappa / \kappa| = 0, \pi/4$ and $\pi/2$.
}
\end{figure}
The left, central and right panels of Fig.~\ref{fig:m_tbarth} show the ${\bar t} t$, $tH$ and $tj$ invariant mass distributions, respectively.
In the ${\bar t} t$ case, the two-body invariant mass distribution is {\it most} peaked in the Standard Model case $\zeta_t = 0$,
whereas for $tH$ and $tj$, the distributions are $least$ peaked in the Standard Model case and $most$ peaked for $|\zeta_t| = \pm \pi/2$, 
as observed in the total invariant mass distribution.

\begin{figure}
\begin{center}
\includegraphics[height=7cm]{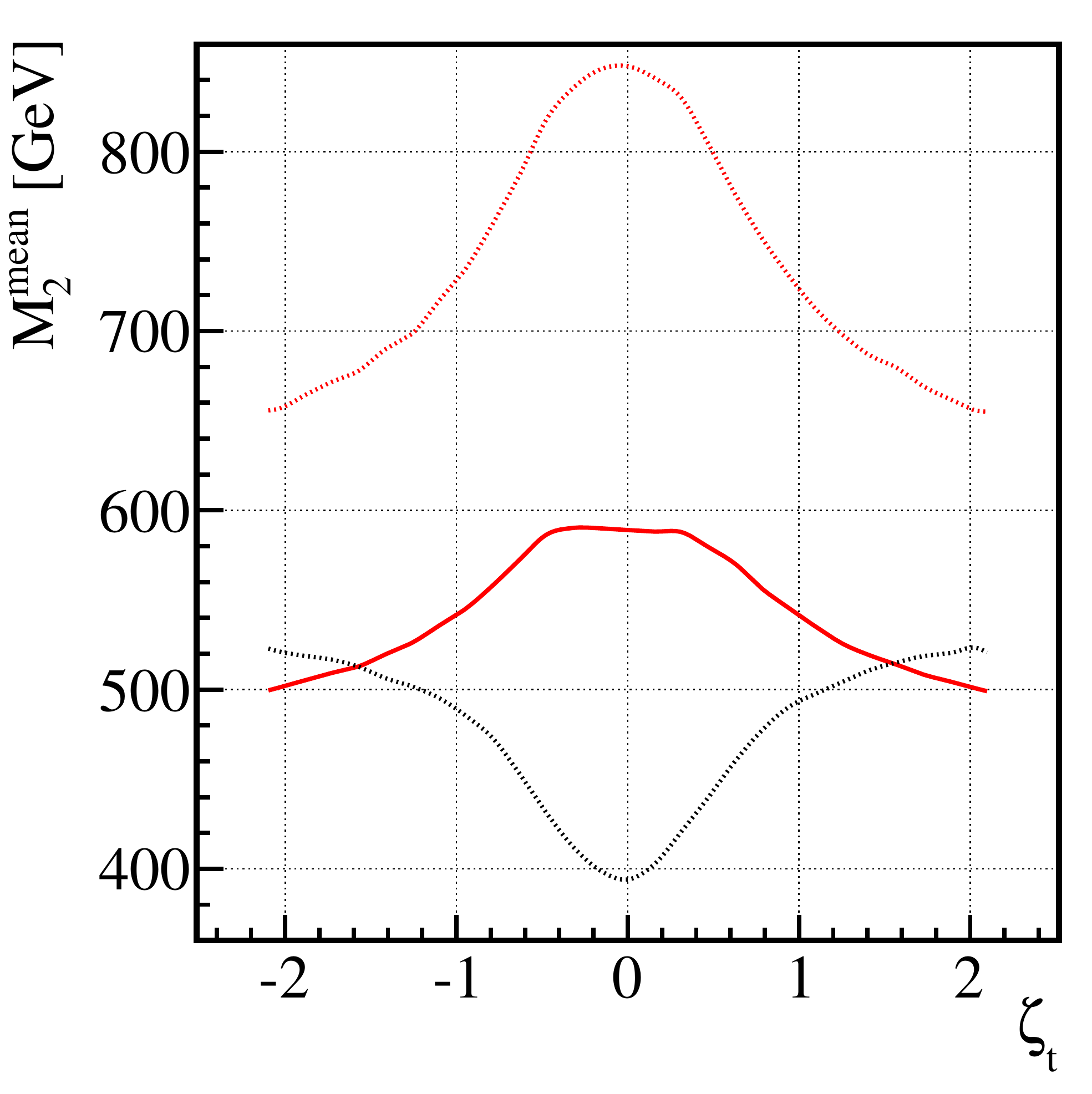}
\hspace{0.5cm}
\includegraphics[height=7cm]{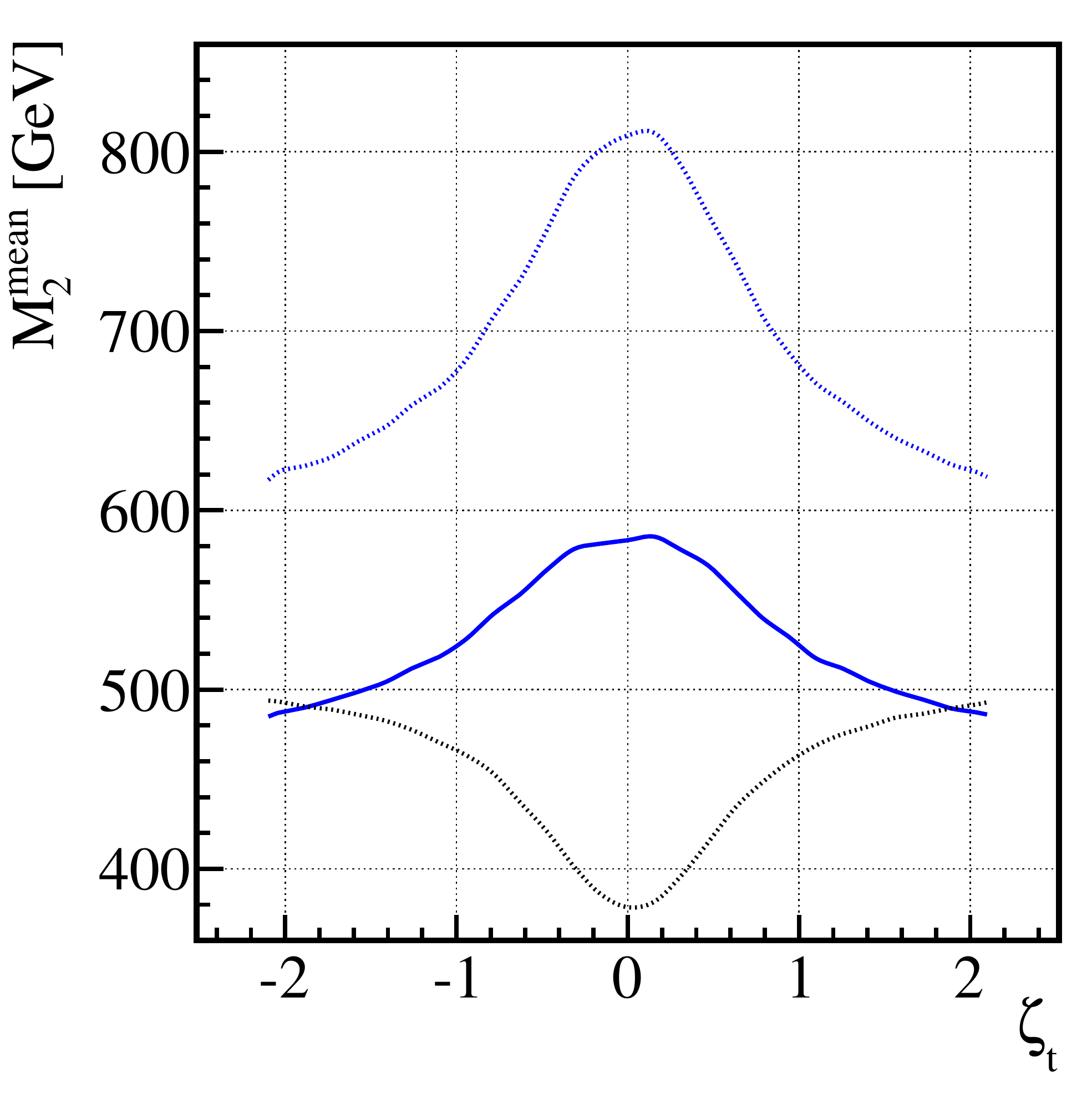}
\end{center}   
\caption{\label{fig:thj_m2body}\it
The mean values for the two-body invariant mass distributions in the $tHj$ and ${\bar t} Hj$ final states (left 
and right panel, respectively) as functions of $\zeta_t$.
The values of $\langle M_{Hj} \rangle$ are indicated by dotted black lines, the values of $\langle M_{tH} \rangle$ and 
$\langle M_{{\bar t}H} \rangle$ are
indicated by solid coloured lines (red and blue, respectively), and the values of $\langle M_{tj} \rangle$ and 
$\langle M_{{\bar t}j} \rangle$ are indicated by dotted coloured lines.}
\end{figure}

The left panel of Fig.~\ref{fig:thj_m2body} shows
the variation with $\zeta_t$ of $\langle M_{H j} \rangle$ (dotted black), $\langle M_{t H} \rangle$ (solid red) and $\langle M_{t j} \rangle$ (dotted red)
along a contour passing trough the middle of the crescent-shape allowed region.
Here we explicitly see $\langle M_{t H} \rangle$ and $\langle M_{t j} \rangle$ are {\it maximised} for the Standard Model case, whilst $\langle M_{Hj} \rangle$ is {\it minimized} in this case.
Although the threshold of the $tj$ invariant mass is smaller than that for the $t H$ invariant mass, 
$\langle M_{tj} \rangle$ is larger than $\langle M_{t H} \rangle$ and
is indeed the largest among the three two-particle invariant masses 
in the 68\% CL. allowed region. 
In the Standard Model case, $\langle M_{Hj} \rangle$ is smaller than $\langle M_{tH} \rangle$.
However this relation becomes reversed near the two tips of the crescent-shape allowed 
region because of the increase and decrease
in $\langle M_{Hj} \rangle$ and $\langle M_{tH} \rangle$, respectively, as $\zeta_t$ deviates from the Standard Model value.   

We also show 
the variation with $\zeta_t$ of $\langle M_{H j} \rangle$ (dotted black), $\langle M_{\bar t H} \rangle$ (solid blue) and $\langle M_{\bar t j} \rangle$ (dotted blue) in $\bar t H j$ production events in the right panel of Fig.~\ref{fig:thj_m2body}.
As can be seen, the features we have discussed above for the $t H j$ events are also found for $\bar t H j$ production events. 
However, we also note a tendency for the two-particle invariant masses in $\bar t H j$ production events
to be somewhat smaller than the corresponding invariant masses in $t H j$ production events. This can be traced
back to the different initial-state parton distributions involved in $t H j$ and $\bar t H j$ production.

\section{Top Polarization Measurements}

We now consider the additional information on the top-$H$ couplings that
could be obtained from measurements of the top-(anti)quark polarization(s).
In principle, there are two classes of measurements: single-spin measurements
in ${\bar t} t H$ and single $tH$ (${\bar t} H$) production with an accompanying light-quark jet,
and measurements of spin-correlations in ${\bar t} t H$ production. Further, one
can measure the single-(anti)top polarization either in the production plane or
perpendicular to it. The latter is particularly interesting, as it violates CP at the tree level.

\subsection{Single-Spin Measurements}

It is easy to see that the single-spin asymmetries actually vanish in ${\bar t} t H$ production, because
of the Dirac matrix factors in the vertices.
However, the single-spin measurements are interesting for $t Hj$ and $\bar t Hj$ production,
because of the $1 - \gamma_5$ factor in the $Wtb$ coupling. As already noted,
the matrix elements of these processes have two competing Feynman diagrams: one is proportional to the $\bar t t H$ coupling
and the other to the $WWH$ coupling, as seen in the lower panel of Fig.~\ref{fig:diagrams}.  
In the latter diagram, the $t$ (or $\bar t$) is emitted from the initial $b$($\bar b$)-quark when it exchanges a 
$W$ boson with a quark (or antiquark) in the other proton.
This $t$($\bar t$) quark therefore prefers the left-handed chirality.  
In the former diagram, $t$ (or $\bar t$) is produced in the same way but subsequently emits a $H$, changing its chirality.
One can therefore expect that the tops in these processes are polarized to some extent,
depending on the details of the $\bar t t H$ coupling.   
 
The angular distributions of the top decay products are correlated with the top spin direction in the following 
way~\cite{Jezabek:1994qs, Brandenburg:2002xr, Godbole:2010kr}:
\begin{equation}
\frac{1}{\Gamma_f} \frac{d \Gamma_f}{d \cos\theta_f}=\frac{1}{2} (1 + \omega_f P_t \cos \theta_f)\, ,
\end{equation}
where $f$ is the type of top decay product: $f = b, \ell, ...$, $\theta_f$ is the angle between the decay 
product $f$ and the top spin quantization axis measured in the rest frame of the top, and $P_t$ is the degree of the top polarization:  
\begin{equation}
P_t = \frac{N(\uparrow) - N(\downarrow)}{N(\uparrow) + N(\downarrow)}\,  .
\end{equation}
The coefficient $\omega_f$ depends on the type of decay product, e.g., $\omega_W = - \omega_b = 0.41$ and $\omega_\ell = 1$ at tree level.

We consider first the angle $\theta_{\ell}$
between the direction of the $t$ and the final-state lepton $\ell$ measured at the rest frame of the top in $tHj$ production events. 
\begin{figure}
\begin{center}
\includegraphics[height=7cm]{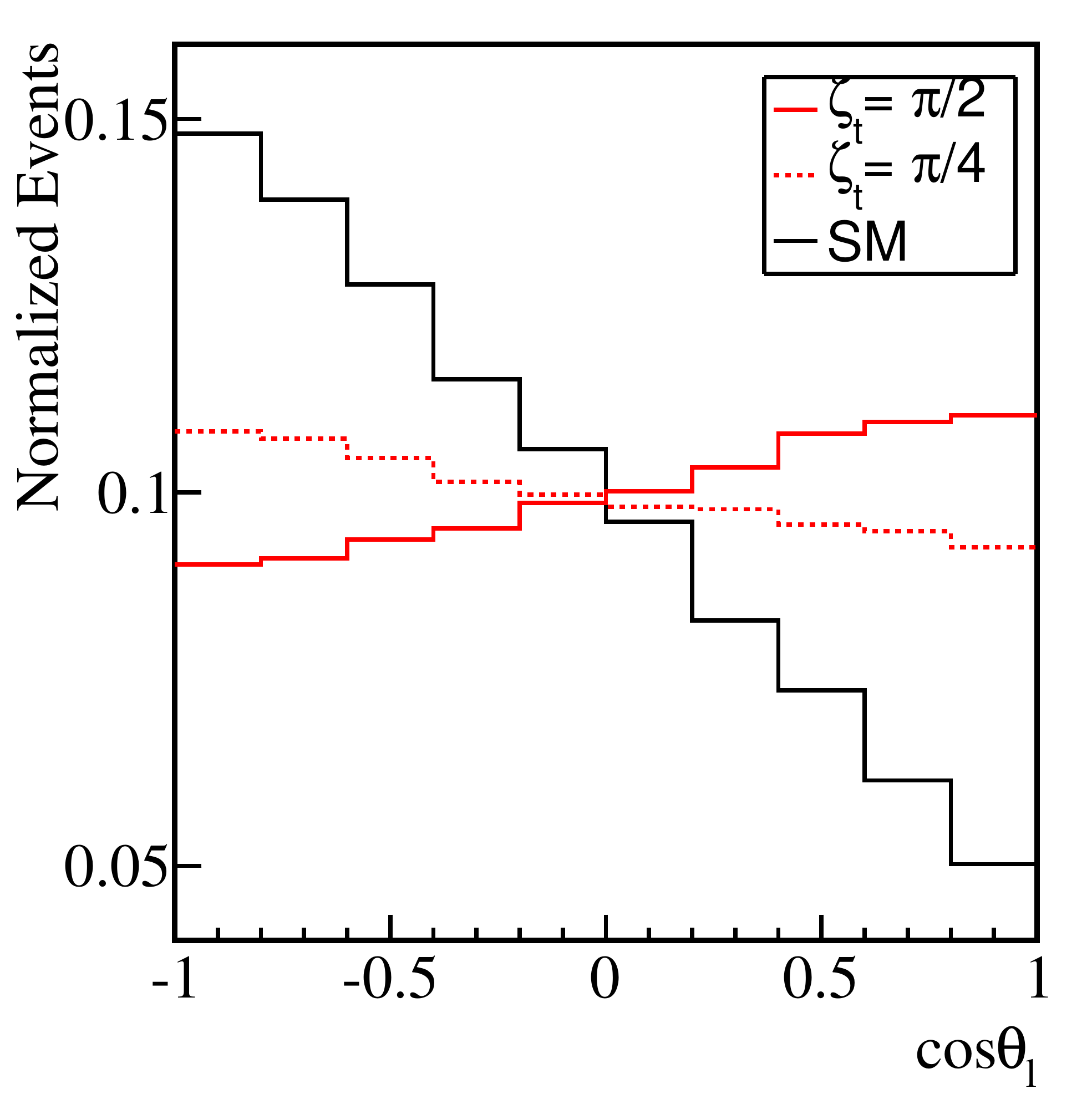} 
\hspace{0.5cm}
\includegraphics[height=7cm]{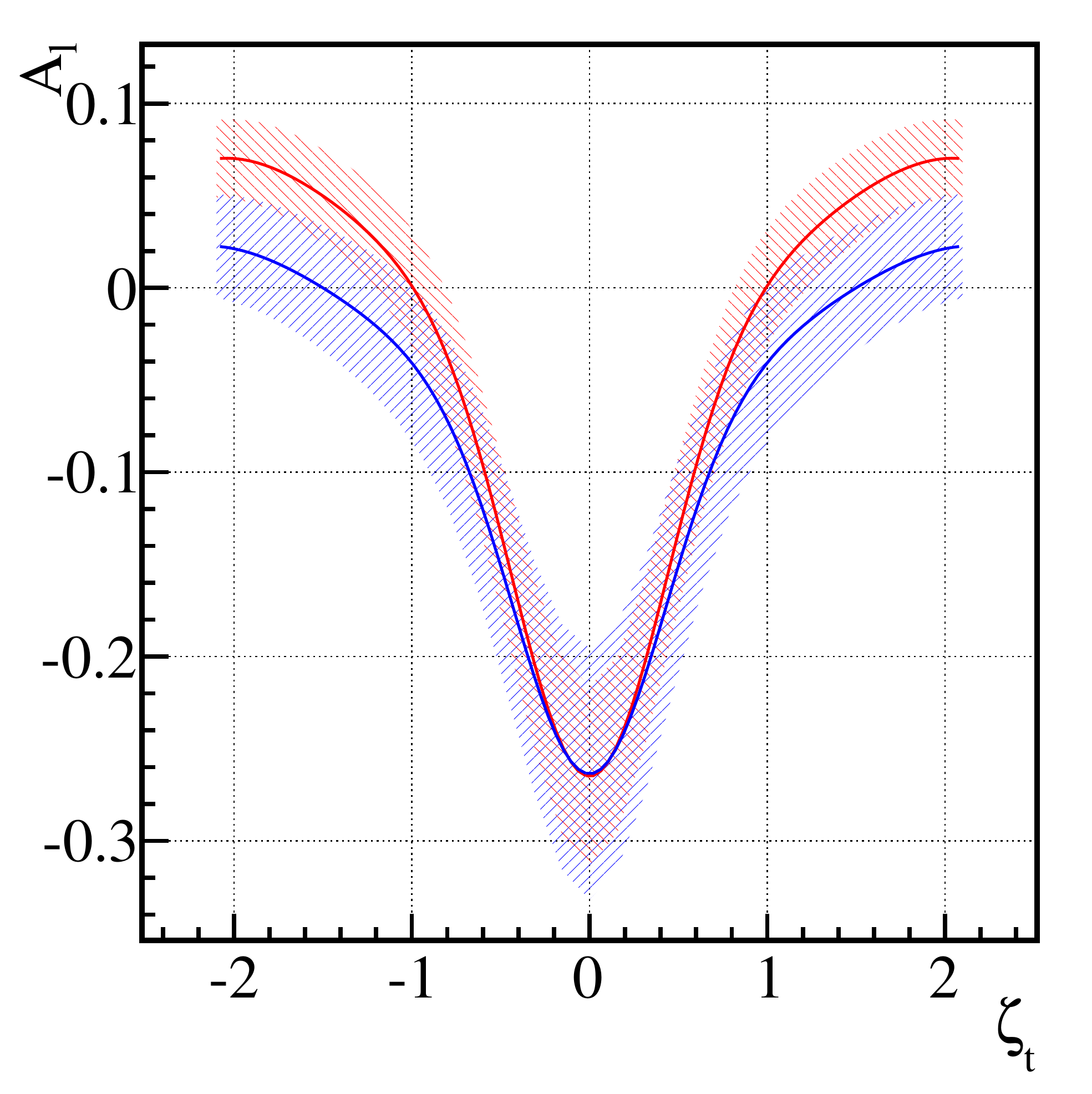} 
\end{center}   
\caption{\label{fig:parallel}\it
Left panel: The distributions in the semileptonic decay angle $\theta_{\ell}$ for the $tHj$ final state
for the indicated values of $\zeta_t$. In the right panel we display the
variation of the forward-backward asymmetry in $\theta_{\ell }$, $A_l$,
with $\zeta_t$ for $tHj$ (${\bar t} H j$) production in red (blue): the shading represents an 
estimate of the measurement error with 100/fb
of integrated luminosity at 14~TeV.
}
\end{figure}
The left panel of Fig.~\ref{fig:parallel} displays the $\cos \theta_{\ell}$ distributions.
As previously, the distribution
for the Standard Model case $\zeta_t = 0$ is shown in black, 
and the distributions for $|\zeta_t| = \pi/4$
and $\pi/2$ in dotted and solid red, respectively.
We can see that the lepton momentum in the Standard Model case strongly prefers the opposite direction to the top's boost direction 
at the top's rest frame, meaning that tops are negatively polarized, $P_t <0$.
As $|\zeta_t|$ increases this preference is weakend.   
For $|\zeta_t| = \pi/4$ the distribution is already quite flat, and the slope is even positive, $P_t \ga 0$, for $|\zeta_t| = \pi/2$. 

The dependence on $\zeta_t$ can more explicitly be seen in the right panel of Fig.~\ref{fig:parallel}, which displays 
the variation with $\zeta_t$ of the forward-backward asymmetry
\begin{equation}
A_{\ell } = \frac{N(\cos \theta_{\ell } > 0) - N(\cos \theta_{\ell } < 0)}{N(\cos \theta_{\ell } > 0) + N(\cos \theta_{\ell } < 0) } ~,
\label{asymdef}
\end{equation}
along a contour passing trough the middle of the crescent-shape allowed region.
The red and blue curves correspond to the $A_{\ell}^t$ and $A_{\ell}^{\bar t}$ in the $t H j$ and $\bar t H j$ production events, respectively.
The shaded bands represent estimates of the measurement error with 100/fb of integrated luminosity at 14~TeV,
again ignoring effects of parton showering, top reconstruction, detector resolution,\footnote{
For studies including these effects, see e.g.~\cite{Buckley:2013lpa, Buckley:2013auc, Papaefstathiou:2011kd}.
} etc..
We see that, within the range of $\zeta_t$ allowed
by the present data, the asymmetry is largest in magnitude (and negative) for $\zeta_t = 0$
(the Standard Model case), is reduced in magnitude for $\zeta_t \ne 0$,
and changes sign for $\zeta_t = \pm \pi/2$. 
On the other hand, there is no sensitivity to the sign of $\zeta_t$.
In the Standard Model case, the asymmetries for the $t H j$ and $\bar t H j$ events are identical.
For $\zeta_t \ne 0$, tops are more positively polarized in the $t H j$ events than in the $\bar t H j$ events. 
 
\begin{figure}
\begin{center}
\includegraphics[height=7cm]{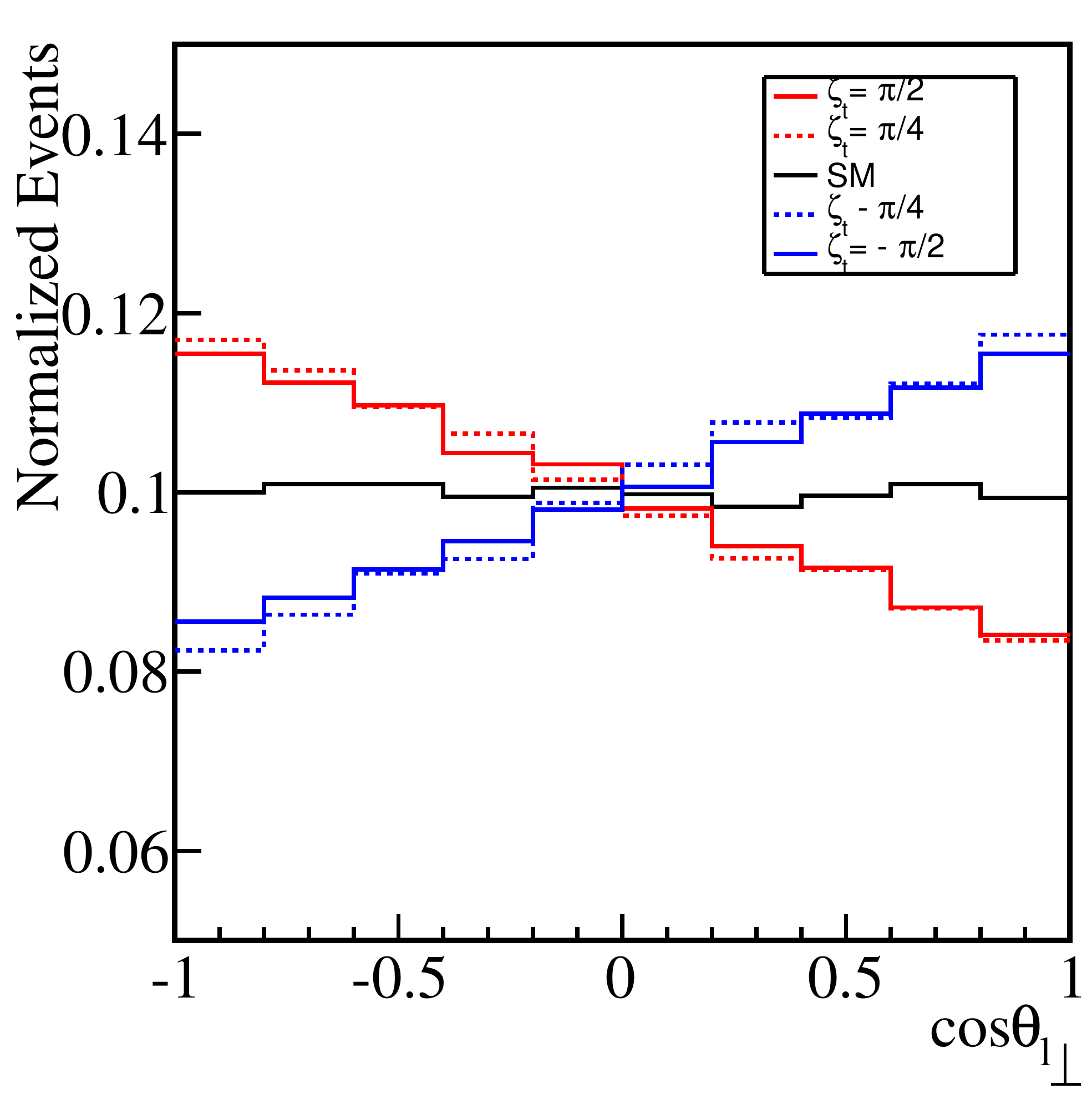} 
\hspace{1cm}
\includegraphics[height=7cm]{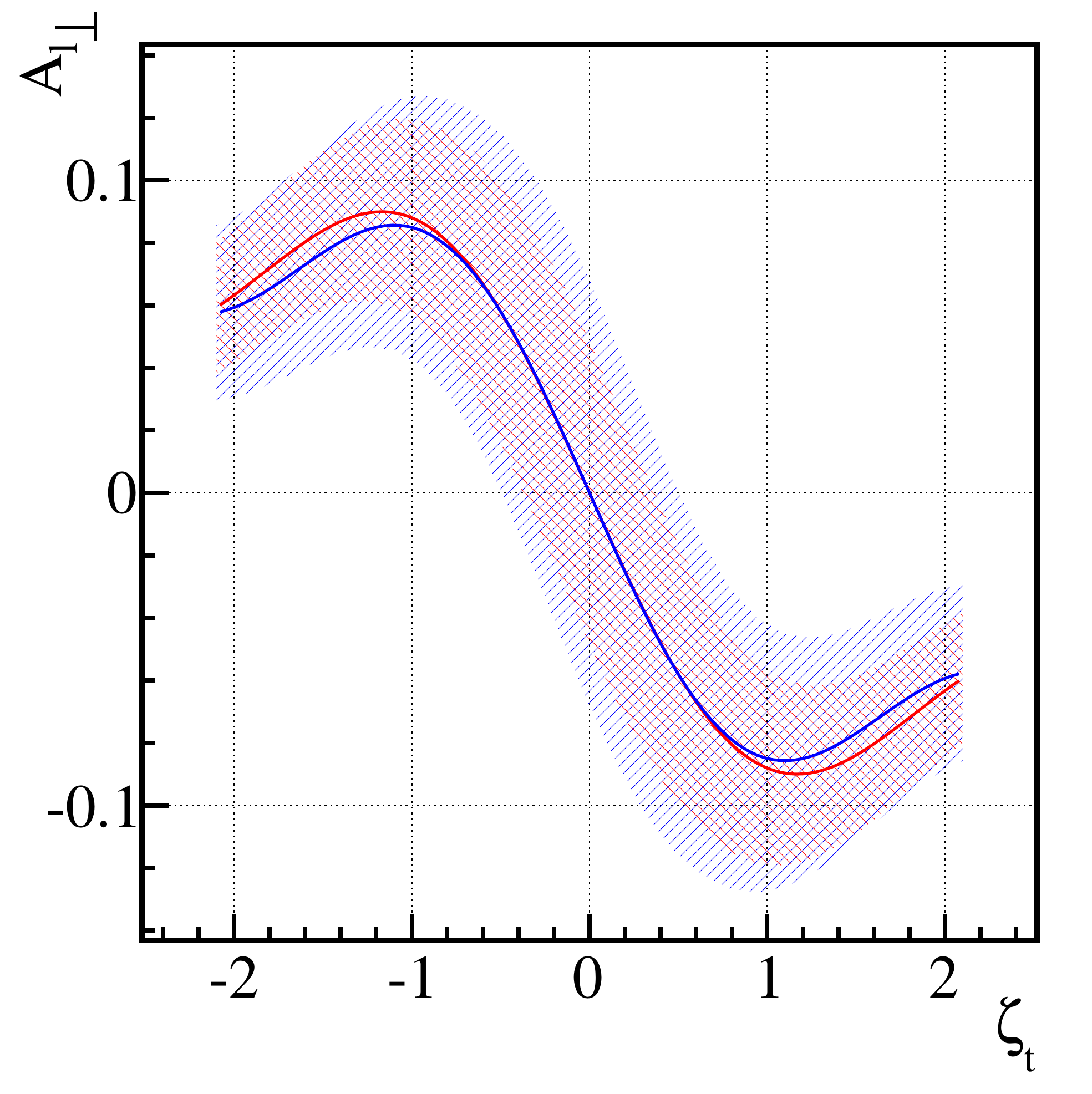} \\
\end{center}   
\caption{\label{fig:perp}\it
Left panel: The distributions in the semileptonic decay angle $\theta_{\ell \perp}$ out of the $tHj$ plane 
for $\zeta_t = {\rm arc} \tan ({\tilde \kappa}_t/ \kappa_t) = 0$ (in black), $\pm \pi/4$
(in dotted red and blue) and $\pm \pi/2$ (in solid red and blue).
Right panel: the asymmetry perpendicular to the plane of $tHj$ (${\bar t} H j$) production, $A_{l \perp}$, as a function of
$\zeta_t$ is indicated in red (blue): the shading represents an estimate of the measurement error with 100/fb
of integrated luminosity at 14~TeV.
}
\end{figure}

We now consider the top (anti-top) polarization perpendicular to the three-body production plane.
We define the spin quantisation axis by $\overrightarrow p_j \times \overrightarrow p_H$
at the rest frame of the top (anti-top), where $j$ is the forward
jet produced by the final-state quark after radiating a virtual $W$ in the diagrams in
the lower panel of Fig.~\ref{fig:diagrams}.
The left panel of Fig.~\ref{fig:perp} shows the $\cos \theta_{\ell \perp}$ distribution, where $\theta_{\ell \perp}$
is the angle between the lepton momentum and the spin quantization axis defined above at the rest frame of the top.
We see that the distribution is flat for the Standard Model case $\zeta_t = 0$.
One the other hand, when $\zeta_t \ne 0$, the lepton prefers one side of the hemisphere with 
respect to the three-body production plane at the rest frame of the top.
The right panel in Fig.~\ref{fig:perp} shows the variation with $\zeta_t$ of the asymmetry $A_{\ell \perp}$,
which is defined in the same way as in Eq. (\ref{asymdef}) for the $\cos \theta_{\ell \perp}$, 
with the same colour-coding as in Fig.~\ref{fig:parallel}.
As expected, there is no up-down 
asymmetry for the Standard Model case $\zeta_t = 0$, but there is a
measurable asymmetry for $\zeta_t = \pm \pi/4$ and $\pm \pi/2$.
In particular, the sign of the perpendicular asymmetry is sensitive to the sign of
$\zeta_t = {\rm arc} \tan ({\tilde \kappa}_t/ \kappa_t)$. This measurement could
therefore provide a direct probe of CP violation in the top-$H$ couplings.

\subsection{Spin Correlation Measurements}

\begin{figure}
\begin{center}
\includegraphics[height=7cm]{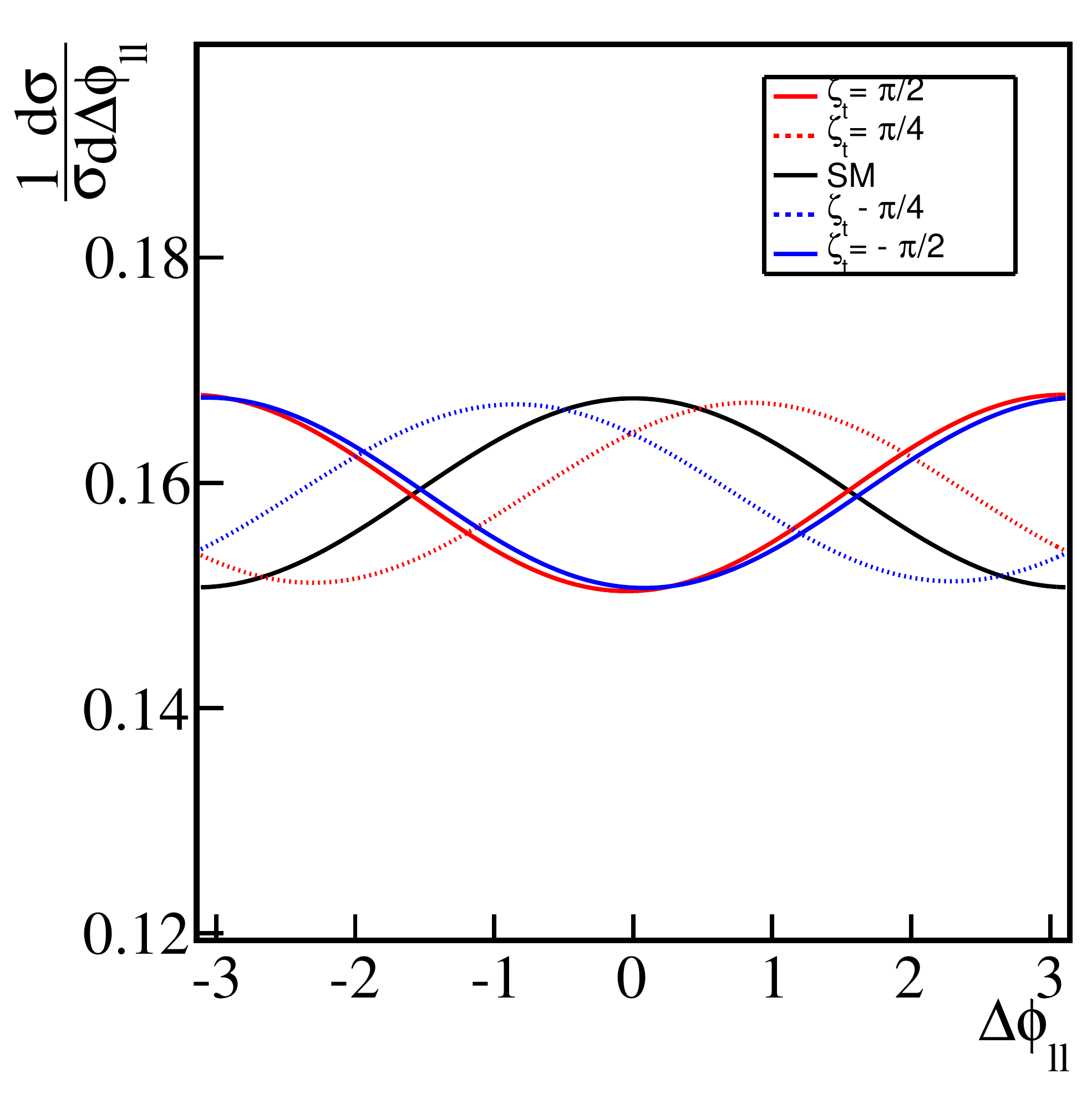}
\hspace{0.5cm}
\includegraphics[height=7cm]{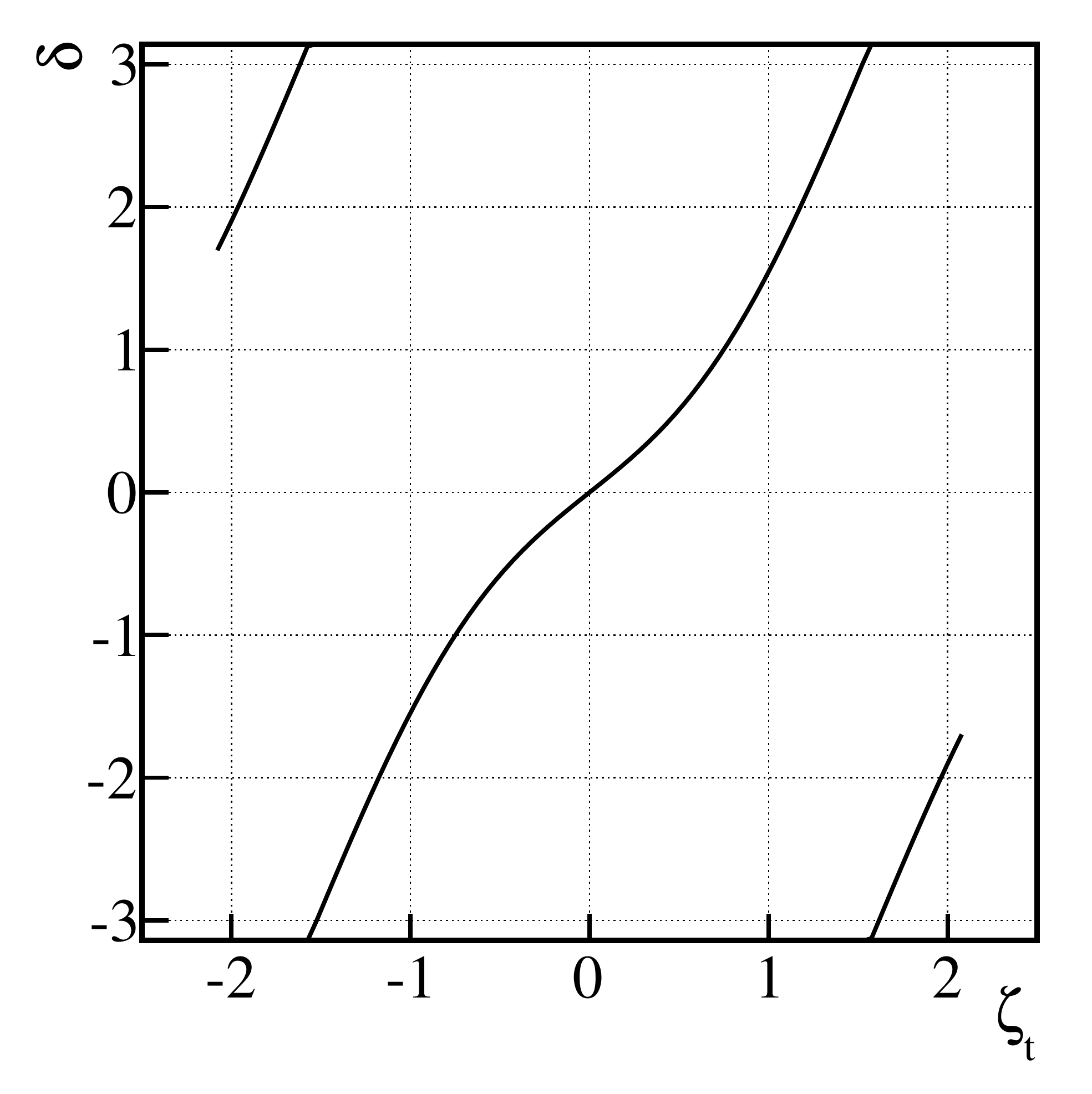}
\end{center}   
\caption{\label{fig:delphi}\it
Left panel: The distributions in the angle $\Delta \phi_{\ell^+ \ell^-}$ between the leptons
produced in $t$ and ${\bar t}$ decay in ${\bar t} tH$ production, in the centre-of-mass of the ${\bar t} t$
system. We display the distributions
for $\zeta_t = {\rm arc} \tan ({\tilde \kappa}_t/ \kappa_t) = 0$ (in black), $\pm \pi/4$
(in dotted red and blue) and $\pm \pi/2$ (in solid red and blue).
Right panel: the phase shift
$\delta$ as a function of $\zeta_t$.
}
\end{figure}

We consider finally possible measurements of the ${\bar t} t$ spin correlation
in ${\bar t}tH$ production. The left panel of Fig.~\ref{fig:delphi} shows the
distribution in the angle $\Delta \phi_{\ell^+ \ell^-}$ between the two lepton momenta
projected onto the plane perpendicular to the $t$ direction at the centre-of-mass frame of the ${\bar t} t$ system. 
The sign of $\Delta \phi_{\ell^+ \ell^-}$ is defined as 
the sign of $\overrightarrow p_t \cdot (\overrightarrow p_{\ell^-} \times \overrightarrow p_{\ell^+})$.\footnote{
The $\Delta \phi_{\ell^+ \ell^-}$ variable is commonly used in the spin correlation measurement in the $\bar t t$ process~\cite{ATLAS:2012ao, CMS:2012gba},
although $\Delta \phi_{\ell^+ \ell^-}$ is defined at the lab frame and its range is [0, $\pi$].
In order to identify CP violation, it is crucial to measure $\Delta \phi_{\ell^+ \ell^-}$ with respect to the top (or anti-top) axis in the range of [$-\pi$, $\pi$].   
}
As previously, the distribution
for the Standard Model case $\zeta_t = {\rm arc} \tan ({\tilde \kappa}_t/ \kappa_t) = 0$ is shown in black, 
those for $\zeta_t = \pm \pi/4$ as dotted lines, and those
for $\pm \pi/2$ as solid lines  (red and blue for $\zeta_t >, < 0$, respectively). 
We see that the distribution has the form
\begin{equation}
\frac{d \sigma}{d \Delta \phi_{\ell^+ \ell^- }} \propto \cos ( \Delta \phi_{\ell^+ \ell^-} - \delta ) + {\rm const.}
\end{equation}
We see in the left panel of Fig.~\ref{fig:delphi} that the phase shift $\delta$ vanishes for the Standard Model case $\zeta_t = 0$,
but takes non-zero values 
for $\zeta_t \ne 0$, and we note that this phase shift is sensitive to the sign of $\zeta_t$. 
The right panel in Fig.~\ref{fig:delphi} shows the value of $\delta$ as a function of $\zeta_t$.
One can see that the $\delta$ varies from $-\pi$ to $\pi$ as $\zeta_t$ varies from $-\pi/2$ to $\pi/2$.
We find that the dependence of $\delta$ on $\zeta_t$ can be very well fitted by the function $\delta = 2 \zeta_t - \sin(2 \zeta_t)/2$.

\section{Summary}

We have shown in this paper that the cross sections and final-state distributions
in ${\bar t} tH$, $t H$ and ${\tilde t} H$ production are sensitive to the 
ratio between the scalar and pseudoscalar top-$H$ couplings $\kappa_t$
and ${\tilde \kappa_t}$.
In particular, the total cross section for ${\bar t} tH$ production {\it decreases}
significantly as the ratio ${\tilde \kappa_t}/\kappa_t$ {\it increases} within
the ranges of values of these couplings that are
allowed by present data on the $H gg$ and $H \gamma \gamma$ couplings.
On the other hand, the total cross sections for $t H$ and ${\tilde t} H$ production
{\it increase} as the ratio ${\tilde \kappa_t}/\kappa_t$ {\it increases}.

We have also found that the invariant mass distributions for the three-body
combinations ${\bar t} tH$, $t Hj$ and ${\tilde t} Hj$ are sensitive to
the ratio ${\tilde \kappa_t}/\kappa_t$, becoming {\it less} peaked at
small masses in the ${\bar t} tH$ case and {\it more} peaked in the 
$t Hj$ and ${\tilde t} Hj$ cases as the ratio ${\tilde \kappa_t}/\kappa_t$ increases.
The two-body invariant mass distributions also carry information about the top-$H$ couplings.

Supplementary information on the ratio ${\tilde \kappa_t}/\kappa_t$ could
be provided by angular distributions in semileptonic $t$ and ${\bar t}$ decays.
In particular, lepton decay angles from the top boost direction could provide information on
the {\it magnitude} of ${\tilde \kappa_t}/\kappa_t$, and lepton decay angles against the $tHj$ (or ${\bar t}Hj$) 
production plane provide information on the {\it sign} of ${\tilde \kappa_t}/\kappa_t$.
Information both on the {\it magnitude}  and {\it sign} of ${\tilde \kappa_t}/\kappa_t$ could also be provided
by measurements of the angle $\Delta \phi_{\ell^+ \ell^-}$ between the directions of
leptons produced in ${\bar t}$ and $t$ decays in the case of ${\bar t} tH$ production.

We conclude that there are good prospects for disentangling the scalar and
pseudoscalar top-$H$ couplings at the LHC via a combination of measurements
of ${\bar t} tH$, $tH$ and ${\bar t}H$ production.

\vspace{1cm}
\noindent{ {\bf Acknowledgments} } \\
\noindent
The work of J.E. and K.S. was supported in part by
the London Centre for Terauniverse Studies (LCTS), using funding from
the European Research Council 
via the Advanced Investigator Grant 267352.
J.E. and M.T. are grateful for funding from the Science and Technology Facilities Council (STFC).
The work of D.S.H. was supported in part by the
Korea Foundation for International Cooperation of Science \&
Technology (KICOS) and the Basic Science Research Programme through
the National Research Foundation of Korea (2013028705).

\end{document}